\documentclass[sigconf]{acmart}

\usepackage{booktabs} % For formal tables

\setcopyright{rightsretained}
\acmDOI{10.1145/3055399.3055413}
\acmISBN{978-1-4503-4528-6/17/06}
\acmConference[STOC'17]{49th Annual ACM SIGACT Symposium on the Theory of Computing}{June 2017}{Montreal, Canada}
\acmYear{2017}
\copyrightyear{2017}
\acmPrice{15.00}

\usepackage[utf8]{inputenc}
\usepackage{amsmath}
\usepackage{amsthm}
\usepackage{amsfonts}
\usepackage{amsopn}
\usepackage{amssymb}
\usepackage{esint}
\usepackage{amsopn}
\usepackage{amssymb}
\usepackage{comment}
\usepackage{appendix}

\usepackage{stmaryrd,mathrsfs}

\usepackage{upgreek}
\DeclareMathOperator{\Lip}{Lip}

\newcommand{\cc}{{\mathsf{c}}}

\newcommand{\N}{\mathbb{N}}

\newcommand{\R}{\mathbb{R}}
\newcommand{\Z}{\mathbb{Z}}
\renewcommand{\H}{\mathbb{H}}

\newcommand{\cH}{\mathcal{H}}

\newcommand{\symdiff}{\mathop{\triangle}}

\newcommand{\vv}{\mathsf{v}}
\newcommand{\hh}{\mathsf{h}}

\renewcommand{\subset}{\subseteq}
\renewcommand{\gamma}{\upgamma}
\renewcommand{\beta}{\upbeta}
\renewcommand{\alpha}{\upalpha}
\renewcommand{\kappa}{\upkappa}
\renewcommand{\psi}{\uppsi}
\renewcommand{\rho}{\uprho}
\renewcommand{\delta}{\updelta}
\renewcommand{\pi}{\uppi}
\renewcommand{\omega}{\upomega}
\renewcommand{\sigma}{\upsigma}
\renewcommand{\emptyset}{\varnothing}
\newcommand{\e}{\varepsilon}
\renewcommand{\theta}{\uptheta}

\newcommand{\ud}[0]{\,\mathrm{d}}
\renewcommand{\epsilon}{\varepsilon}

\renewcommand{\xi}{\upxi}
\renewcommand{\tau}{\uptau}

\newtheorem{thm}{Theorem}[section]

\newtheorem{cor}[thm]{Corollary}

\newtheorem{defn}[thm]{Definition}
\theoremstyle{remark}

\renewcommand{\le}{\leqslant}
\renewcommand{\ge}{\geqslant}

\renewcommand{\setminus}{\smallsetminus}
\newcommand{\eqdef}{\stackrel{\mathrm{def}}{=}}
\newcommand{\1}{\mathbf 1}

\newcommand{\n}{\{1,\ldots,n\}}

\renewcommand{\H}{\mathbb{H}}
\newcommand{\GL}{\mathsf{GL}}
\newcommand{\CH}{\mathcal{H}}

\renewcommand{\gamma}{\upgamma}
\renewcommand{\beta}{\upbeta}
\renewcommand{\alpha}{\upalpha}
\renewcommand{\kappa}{\upkappa}
\renewcommand{\psi}{\uppsi}
\renewcommand{\rho}{\uprho}
\renewcommand{\delta}{\updelta}
\renewcommand{\pi}{\uppi}
\renewcommand{\omega}{\upomega}
\renewcommand{\lambda}{\uplambda}
\renewcommand{\eta}{\upeta}
\renewcommand{\chi}{\upchi}
\renewcommand{\nu}{\upnu}
\renewcommand{\mu}{\upmu}
\renewcommand{\phi}{\upphi}

\newcommand{\aaa}{\mathsf{a}}
\newcommand{\bb}{\mathsf{b}}
\newcommand{\dd}{\mathsf{d}}
\newcommand{\ee}{\mathsf{e}}

\newcommand{\0}{\mathbf{0}}

\begin{document}

\title[The integrality gap of the Goemans--Linial SDP]{The integrality gap of the Goemans--Linial SDP relaxation for Sparsest Cut is at least a constant multiple of $\sqrt{\log n}$}
\titlenote{The research presented here was conducted under the auspices of the Simons Algorithms and Geometry (A\&G) Think Tank. A full version of this extended abstract, titled ``Vertical perimeter versus horizontal perimeter,'' that contains complete proofs and additional results is available at \url{https://arxiv.org/abs/1701.00620}. Nevertheless, this extended abstract contains material that is not included in the full version.}

\author{Assaf Naor}
\authornote{Supported by BSF grant 2010021, the Packard Foundation and the Simons Foundation.}
\affiliation{%
  \institution{Princeton University}
  %\streetaddress{Fine Hall, Washington Road}
  \city{Princeton}
  \state{NJ}
  \country{USA}
  \postcode{08544-1000}
}
\email{naor@math.princeton.edu}

\author{Robert Young}
\authornote{Supported by the NSF and the Sloan Foundation.  This material is based upon work supported by the National Science Foundation under Grant No.\ 1612061.}
\affiliation{%
  \institution{New York University}
  %\streetaddress{251 Mercer Street}
  \city{New York}
  \state{NY}
  \country{USA}
  \postcode{10012-1185}
}
\email{ryoung@cims.nyu.edu}

\begin{abstract}
We prove that the integrality gap of the Goemans--Linial semidefinite programming relaxation for the Sparsest Cut Problem is $\Omega(\sqrt{\log n})$ on inputs with $n$ vertices, thus matching the upper bound $(\log n)^{\frac12+o(1)}$ of~\cite{ALN08} up to lower-order factors. This statement is a consequence of the following new isoperimetric-type  inequality. Consider the $8$-regular graph whose vertex set is the $5$-dimensional integer grid $\Z^5$ and where each vertex $(\mathsf{a},\mathsf{b},\mathsf{c},\mathsf{d},\mathsf{e})\in \Z^5$ is connected to the $8$ vertices
$(\mathsf{a}\pm 1,\mathsf{b},\mathsf{c},\mathsf{d},\mathsf{e})$,   $(\mathsf{a},\mathsf{b}\pm 1,\mathsf{c},\mathsf{d},\mathsf{e})$,  $(\mathsf{a},\mathsf{b},\mathsf{c}\pm 1,\mathsf{d},\mathsf{e}\pm \mathsf{a})$, $(\mathsf{a},\mathsf{b},\mathsf{c},\mathsf{d}\pm 1,\mathsf{e}\pm \mathsf{b})$.
This graph is known as the Cayley graph of the 5-dimensional discrete Heisenberg group. Given $\Omega\subset \Z^5$, denote the size of its edge boundary in this graph (a.k.a.\ the {\em horizontal perimeter} of $\Omega$) by $|\partial_{\hh}\Omega|$. For $t\in \N$, denote by  $|\partial^t_{\vv}\Omega|$ the number of $(\mathsf{a},\mathsf{b},\mathsf{c},\mathsf{d},\mathsf{e})\in \Z^5$ such that exactly one of the two vectors $(\mathsf{a},\mathsf{b},\mathsf{c},\mathsf{d},\mathsf{e}),(\mathsf{a},\mathsf{b},\mathsf{c},\mathsf{d},\mathsf{e}+t)$ is in $\Omega$. The {\em vertical perimeter}  of $\Omega$ is defined to be $|\partial_{\vv}\Omega|= \sqrt{\sum_{t=1}^\infty |\partial^t_{\vv}\Omega|^2/t^2}$. We show that every subset $\Omega\subset \Z^5$ satisfies $|\partial_{\vv}\Omega|=O(|\partial_{\hh}\Omega|)$.  This {\em vertical-versus-horizontal isoperimetric inequality} yields the above-stated integrality gap for   Sparsest Cut and answers several geometric and analytic questions of independent interest.

The theorem stated above is the culmination of a program that was pursued in the works~\cite{LN06,CK10,CheegerKleinerMetricDiff,CKN09,CKN,AusNaoTes,LafforgueNaor}  whose aim is to understand the performance of the Goemans--Linial semidefinite program through the embeddability properties of Heisenberg groups. These investigations have mathematical significance even beyond their established relevance to approximation algorithms and combinatorial optimization.  In particular they contribute to a range of mathematical disciplines including functional analysis, geometric group theory, harmonic analysis, sub-Riemannian geometry, geometric measure theory, ergodic theory, group representations, and metric differentiation.  This article builds on the above cited works, with the ``twist'' that while those works were equally valid for any finite dimensional Heisenberg group, our result holds  for the Heisenberg group of dimension $5$ (or higher) but {\em fails} for the $3$-dimensional Heisenberg group. This insight leads to our core contribution, which is a deduction of an endpoint $L_1$-boundedness of a certain singular integral on $\R^5$ from the (local)  $L_2$-boundedness of the corresponding singular integral on $\R^3$.  To do this, we devise  a corona-type decomposition of subsets of a Heisenberg group, in the spirit of the construction~\cite{DSAnalysis} that David and Semmes performed in $\R^n$, but with two main conceptual differences (in addition to more technical differences that arise from the peculiarities of the geometry of Heisenberg group).  Firstly, the``atoms'' of our decomposition are perturbations of {\em intrinsic Lipschitz graphs} in the sense of Franchi, Serapioni, and Serra Cassano~\cite{FSSC06} (plus the requisite ``wild'' regions that satisfy a Carleson packing condition).  Secondly, we control the local overlap   of our corona decomposition by using quantitative monotonicity rather than Jones-type $\beta$-numbers.
\end{abstract}



%\date{\today}

\maketitle

% \tableofcontents
\section{Introduction}

Fix $n\in \N$. The input of the {\em Sparsest Cut Problem} consists of two $n$ by $n$ symmetric matrices with nonnegative entries $C=(C_{ij}),D=(D_{ij})\in M_n([0,\infty))$, which are commonly called capacities and demands, respectively. The goal is to design a polynomial-time algorithm to evaluate the quantity
\begin{equation}\label{eq:def opt}
\mathsf{OPT}(C,D)\eqdef \min_{\emptyset\subsetneq A\subsetneq \n}\frac{\sum_{(i,j)\in A\times (\n\setminus A)}C_{ij}}{\sum_{(i,j)\in A\times (\n\setminus A)}D_{ij}}.
\end{equation}

In view of the extensive literature on the Sparsest Cut Problem, it would be needlessly repetitive to recount here the rich and multifaceted impact of this optimization problem on computer science and mathematics; see instead the articles~\cite{AKRR90,LR99}, the surveys~\cite{Shm95,Lin02,Chawla08,Nao10}, Chapter~10 of the monograph~\cite{DL97}, Chapter~15 of the monograph~\cite{Mat02}, Chapter~1 of the monograph~\cite{Ost13}, and the references therein. It suffices to say  that by tuning the choice of matrices $C,D$ to the  problem at hand, the  minimization in~\eqref{eq:def opt} finds a partition of the ``universe'' $\n$ into two parts, namely the sets $A$ and $\n\setminus A$, whose appropriately weighted interface is as small as possible, thus allowing for inductive solutions of various algorithmic tasks, a procedure known as {\em divide and conquer}. (Not all of the uses of the  Sparsest Cut Problem fit into this framework. A recent algorithmic application of a different nature can be found in~\cite{MMV14}.)

It is $NP$-hard to compute $\mathsf{OPT}(C,D)$ in polynomial time~\cite{SM90}. By~\cite{CK09-hardness} there exists $\e_0>0$ such that it is even $NP$-hard to compute $\mathsf{OPT}(C,D)$ within a multiplicative factor of less than $1+\e_0$. If one assumes Khot's Unique Games Conjecture~\cite{Kho02,Kho10,Tre12} then by~\cite{CKKRS06,KV15} there does not exist a polynomial-time algorithm that can compute $\mathsf{OPT}(C,D)$ within any universal constant factor.

By the above hardness results, a much more realistic goal would be to design a polynomial-time algorithm that takes as input the capacity and demand matrices $C,D\in M_n([0,\infty))$ and outputs a number $\mathsf{ALG}(C,D)$ that is guaranteed to satisfy $$\mathsf{ALG}(C,D)\le \mathsf{OPT}(C,D)\le \rho(n)\mathsf{ALG}(C,D),$$ with (hopefully) the quantity $\rho(n)$ growing to $\infty$ slowly as $n\to \infty$. Determining the best possible asymptotic behaviour of $\rho(n)$ (assuming $P\neq NP$) is an  open problem of major importance.

In~\cite{LLR95,AR98}  an algorithm was designed, based on linear programming (through the connection to multicommodity flows) and Bourgain's embedding theorem~\cite{Bou85}, which yields $\rho(n)=O(\log n)$. An algorithm based on semidefinite programming (to be described precisely below) was proposed by Goemans and Linial in the mid-1990s.  To the best of our knowledge this idea first appeared in the literature in~\cite[page~158]{Goe97}, where it was speculated that it might yield a constant factor approximation for Sparsest Cut (see also~\cite{Lin02,Lin-open}). In what follows, we  denote the approximation ratio of the Goemans--Linial algorithm on inputs of size at most $n$ by $\rho_{\GL}(n)$. The hope that $\rho_{\GL}(n)=O(1)$ was dashed in the remarkable work~\cite{KV15}, where the lower bound $\rho_{\mathsf{GL}}(n)\gtrsim \sqrt[6]{\log\log n}$ was proven.\footnote{Here, and in what follows, we use the following (standard) asymptotic notation. Given $a,b>0$, the notations
$a\lesssim b$ and $b\gtrsim a$ mean that $a\le \mathsf{K}b$ for some
universal constant $\mathsf{K}>0$. The notation $a\asymp b$
stands for $(a\lesssim b) \wedge  (b\lesssim a)$. Thus $a\lesssim b$ and $a\gtrsim b$ are the same as $a=O(b)$ and $a=\Omega(b)$, respectively, and $a\asymp b$ is the same as $a=\Theta(b)$.} An improved analysis of the ideas of~\cite{KV15} was conducted in~\cite{KR09}, yielding the estimate $\uprho_{\mathsf{GL}}(n)\gtrsim \log\log n$. An entirely different approach based on the geometry of the Heisenberg group was introduced in~\cite{LN06}.  In combination with the important works~\cite{CK10,CheegerKleinerMetricDiff} it gives a different proof that $\lim_{n\to \infty}\rho_{\GL}(n)=\infty$. In~\cite{CKN09,CKN} the previously best-known bound $\rho_{\mathsf{GL}}(n)\gtrsim (\log n)^\delta$ was obtained for an effective (but small) positive universal constant $\delta$.

%Section~\ref{sec:previous} will contain a more detailed overview of the ideas and tools that have been used to obtain these bounds.

 Despite these lower bounds, the Goemans--Linial algorithm yields an approximation ratio of $o(\log n)$, so it is asymptotically more accurate than the linear program of~\cite{LLR95,AR98}.  Specifically, in~\cite{CGR08} it was shown that $\rho_{\GL}(n)\lesssim (\log n)^{\frac34}$.  This was improved in~\cite{ALN08} to $\rho_{\GL}(n)\lesssim (\log n)^{\frac12+o(1)}$. See Section~\ref{sec:previous} below for additional background on the  results quoted above. No other polynomial-time algorithm for the Sparsest Cut problem is known (or conjectured) to have an approximation ratio that is asymptotically better than that of the Goemans--Linial algorithm. However, despite major scrutiny by researchers in approximation algorithms, the asymptotic behavior of $\rho_{\GL}(n)$ as $n\to \infty$ remained unknown. Theorem~\ref{thm:main GL lower intro} below resolves this question up to lower-order factors.

\begin{thm}\label{thm:main GL lower intro}
The approximation ratio of the Goemans--Linial algorithm satisfies $\rho_{\GL}(n)\gtrsim \sqrt{\log n}$.
\end{thm}

\subsection{The SDP relaxation} The Goemans--Linial algorithm is simple to describe. It takes as input the symmetric matrices $C,D\in M_n([0,\infty))$ and proceeds to compute the following quantity.
\begin{equation*}\label{eq:def sdp}
\mathsf{SDP}(C,D)\eqdef \inf_{(v_1,\ldots,v_n)\in \mathsf{NEG}_n} \frac{\sum_{i=1}^n\sum_{j=1}^nC_{ij}\|v_i-v_j\|_{2}^2}{\sum_{i=1}^n\sum_{j=1}^nD_{ij}\|v_i-v_j\|_{2}^2},
\end{equation*}
where
\begin{align*}
\mathsf{NEG}_n&\eqdef \Big\{(v_1,\ldots v_n)\in (\R^n)^n:\\ &\qquad\qquad \|v_i-v_j\|_2^2\le \|v_i-v_k\|_2^2+\|v_k-v_j\|_2^2\\ &\qquad\qquad\mathrm{for\ all\ } i,j,k\in \n\Big\}.
\end{align*}
Thus $\mathsf{NEG}_n$  is the set of $n$-tuples $(v_1,\ldots v_n)$ of vectors in $\R^n$ such that $(\{v_1,\ldots,v_n\},\upnu_n)$ is a semi-metric space, where  $\upnu_n:\R^n\times \R^n\to [0,\infty)$ is defined by  $\upnu_n(x,y)=\sum_{j=1}^n (x_j-y_j)^2=\|x-y\|_2^2$ for every $x=(x_1,\ldots,x_n),y=(y_1,\ldots,y_n)\in \R^n$. A semi-metric space $(X,d_X)$ is said~\cite{DL97} to be of {\em negative type}  if $(X,\sqrt{d_X})$ embeds isometrically into a Hilbert space. So, $\mathsf{NEG}_n$ can be  described as the set of all (ordered) negative type semi-metrics of size $n$. It is simple to check that the evaluation of the quantity $\mathsf{SDP}(C,D)$ can be cast as a semidefinite program (SDP), so it can be achieved (up to $o(1)$ precision) in polynomial time~\cite{GLS93}. One has $\mathsf{SDP}(C,D)\le  \mathsf{OPT}(C,D)$ for all symmetric matrices $C,D\in M_n([0,\infty))$. See  e.g.~\cite[Section~15.9]{Mat02-book} or~\cite[Section~4.3]{Nao10} for an explanation of the above assertions about $\mathsf{SDP}(C,D)$, as well as additional background and motivation. The pertinent question is therefore to evaluate the asymptotic behavior as $n\to \infty$ of the sequence
$$
\uprho_{\mathsf{GL}}(n)\eqdef \sup_{\substack{C,D\in M_n([0,\infty))\\ C,D\ \mathrm{symmetric}}} \frac{\mathsf{OPT}(C,D)}{\mathsf{SDP}(C,D)}.
$$
This is the quantity $\rho_{\GL}(n)$ appearing in Theorem~\ref{thm:main GL lower intro}, also known as the  {\em integrality gap} of the Goemans--Linial semidefinite programming relaxation for the Sparsest Cut Problem.

\subsection{Bi-Lipschitz embeddings}\label{sec:embed intro} A duality argument of Rabinovich (see~\cite[Lemma~4.5]{Nao10} or~\cite[Section~1]{CKN09}) establishes that  $\uprho_{\mathsf{GL}}(n)$ is equal to the largest possible $L_1$-distortion of an $n$-point semi-metric of negative type.  If $d:\n^2\to [0,\infty)$ is a semi-metric, its $L_1$ distortion, denoted $c_1(\n,d)$, is the smallest $D\in [1,\infty)$ for which there are integrable functions\footnote{If one wishes to use finite-dimensional vectors rather than functions then by~\cite{Wit86} there exist $v_1,\ldots, v_n\in \R^{n(n-1)/2}$ such that $\int_0^1|f_i(t)-f_j(t)|\ud t=\|v_i-v_j\|_1=\sum_{k=1}^{n(n-1)/2} |v_{ik}-v_{jk}|$ for every $i,j\in \n$.}
$f_1,\ldots,f_n:[0,1]\to \R$ such that $\int_0^1|f_i(t)-f_j(t)|\ud t\le d(i,j)\le D\int_0^1|f_i(t)-f_j(t)|\ud t$ for every $i,j\in \n$. Rabinovich's duality argument proves that $\uprho_{\mathsf{GL}}(n)$ is equal to the maximum of $c_1(\n,d)$ over all possible semi-metrics $d$ of negative type on $\n$. Hence, Theorem~\ref{thm:main GL lower intro} is equivalent to the assertion that for every $n\in \N$ there exists a metric of negative type $d:\n^2\to [0,\infty)$ for which $c_1(\n,d)\gtrsim \sqrt{\log n}$.

\subsection{A poorly-embeddable metric} The $5$-dimensional discrete Heisenberg group, denoted  $\H_\Z^5$, is the following group of $4$ by $4$ invertible matrices, equipped with the usual matrix multiplication.
\begin{equation}\label{eq:def H5}
\H_\Z^5\eqdef \left\{\begin{pmatrix} 1 & \aaa &\bb& \ee\\
  0 & 1&0 & \cc\\
  0 & 0 & 1 & \dd \\
            0 & 0 & 0& 1
                       \end{pmatrix}: \aaa,\bb,\cc,\dd,\ee\in \Z\right\}\subset \mathrm{GL}_4(\R).
\end{equation}
This group is generated by the symmetric set  $$S\eqdef \{X_1,X_1^{-1},X_2,X_2^{-1},Y_1,Y_1^{-1},Y_2,Y_2^{-1}\},$$ where
\begin{align}\label{eq:def generators}
\begin{split}
X_1&\eqdef \begin{pmatrix} 1 &  1 &0 & 0\\
  0 & 1&0 & 0\\
  0 & 0 & 1 & 0 \\
            0 & 0 & 0& 1
                       \end{pmatrix},\qquad  X_2\eqdef \begin{pmatrix} 1 & 0 & 1 & 0\\
  0 & 1&0 & 0\\
  0 & 0 & 1 & 0 \\
            0 & 0 & 0& 1
                       \end{pmatrix},\\  Y_1&\eqdef \begin{pmatrix} 1 & 0 &0 & 0\\
  0 & 1&0 &  1\\
  0 & 0 & 1 & 0 \\
            0 & 0 & 0& 1
                       \end{pmatrix},\qquad  Y_2\eqdef \begin{pmatrix} 1 & 0 &0& 0\\
  0 & 1&0 & 0\\
  0 & 0 & 1 &  1 \\
            0 & 0 & 0& 1
                       \end{pmatrix}.
\end{split}
\end{align}

For notational convenience we shall identify the matrix in~\eqref{eq:def H5} with the vector $(\aaa,\bb,\cc,\dd,\ee)\in \Z^5$. This yields an identification of $\H_\Z^5$ with the $5$-dimensional integer grid $\Z^5$.  We view $\Z^5$ as a (noncommutative) group equipped with the product that is inherited from matrix multiplication through the above identification, i.e., for every $(\aaa,\bb,\cc,\dd,\ee),(\alpha,\beta,\gamma,\delta,\upepsilon)\in \Z^5$ we set
\begin{multline}\label{eq:Z5 product}
(\aaa,\bb,\cc,\dd,\ee)(\alpha,\beta,\gamma,\delta,\upepsilon)\\\eqdef (\aaa+\alpha,\bb+\beta,\cc+\gamma,\dd+\delta,\ee+\upepsilon+\aaa\gamma+\bb\delta).
\end{multline}

Note that under the above identification the identity element of $\H_\Z^5$ is the zero vector $\mathbf{0}\in \Z^5$, the inverse of an element $h=(\aaa,\bb,\cc,\dd,\ee)\in \Z^5$ is $h^{-1}=(-\aaa,-\bb,-\cc,-\dd,-\ee+\aaa\cc+\bb\dd)$, and the generators $X_1,X_2,Y_1,Y_2$ in~\eqref{eq:def generators} are the first four standard basis elements of $\R^5$. Let $Z$ denote the fifth standard basis element of $\R^5$, i.e., $Z=(0,0,0,0,1)$. We then have the relations $[X_1,Y_1]=[X_2,Y_2]=Z$ and $[X_1,X_2]=[X_1,Y_2]=[X_1,Z]=[Y_1,X_2]=[Y_1,Y_2]=[Y_1,Z]=[X_2,Z]=[Y_2,Z]=\0$, where we recall the standard commutator notation $[g,h]=ghg^{-1}h^{-1}$ for every two group elements $g,h\in \H_\Z^5$. In other words, any two elements from $\{X_1,X_2,Y_1,Y_2,Z\}$ other than $X_1,Y_1$ or $X_2,Y_2$ commute, and the commutators of $X_1,Y_1$ and $X_2,Y_2$ are both equal to $Z$. In particular, $Z$ commutes with all of the members of the generating set $S$, and therefore $Z$ is in the {\em center} of $\H_\Z^5$. It is worthwhile to mention that these commutation relations could be used to define the group $\H_\Z^5$ abstractly using generators and relations, but this fact will not be needed in what follows.

This group structure induces a graph $\mathcal{X}_S(\H_\Z^{5})$ on $\Z^5$, called the \emph{Cayley graph} of $\H^{5}_\Z$.  The edges of this graph are defined to be the unordered pairs of the form $\{h,hs\}$, where $h\in \Z^5$ and $s\in S$. This is an $8$-regular connected graph, and by the group law~\eqref{eq:Z5 product}, the neighbors of each vertex $(\mathsf{a},\mathsf{b},\mathsf{c},\mathsf{d},\mathsf{e})\in \Z^5$ are
$(\mathsf{a}\pm 1,\mathsf{b},\mathsf{c},\mathsf{d},\mathsf{e}),  (\mathsf{a},\mathsf{b}\pm 1,\mathsf{c},\mathsf{d},\mathsf{e}),  (\mathsf{a},\mathsf{b},\mathsf{c}\pm 1,\mathsf{d},\mathsf{e}\pm \mathsf{a}), (\mathsf{a},\mathsf{b},\mathsf{c},\mathsf{d}\pm 1,\mathsf{e}\pm \mathsf{b}).$  The shortest-path metric on $\Z^5$ that is induced by this graph structure will be denoted below by $d_W:\Z^5\times \Z^5\to \N\cup\{0\}$. This metric is also known as the left-invariant \emph{word metric} on the Heisenberg group $\H_\Z^5$. For every $R\in [0,\infty)$ denote the (closed) ball of radius $R$ centered at the identity element by $\mathcal{B}_R=\{h\in \Z^5:\ d_W(h,\0)\le R\}$.  It is well-known (see e.g.~\cite{Bas72}) that $|\mathcal{B}_R|\asymp R^6$ and $d_W(\0,Z^R)\asymp\sqrt{R}$ for every $R\in \N$.  Our main result is the following theorem.

\begin{thm}\label{thm:distortion R}
For all $R\ge 2$ we have $c_1(\mathcal{B}_{R},d_W)\asymp \sqrt{\log R}$.
\end{thm}
The new content of Theorem~\ref{thm:distortion R} is the bound $c_1(\mathcal{B}_R,d_W)\gtrsim \sqrt{\log R}$. The matching upper bound $c_1(\mathcal{B}_R,d_W)\lesssim \sqrt{\log R}$ has several proofs in the literature; see e.g.~the discussion immediately following Corollary~1.3 in~\cite{LafforgueNaor} or Section~\ref{sec:embed intro}  below.  The previous best known estimate~\cite{CKN} was that there exists a universal constant $\updelta>0$ such that  $c_1(\mathcal{B}_R,d_W)\ge (\log R)^\updelta$. By~\cite[Theorem~2.2]{LN06} the metric $d_W$ is bi-Lipschitz equivalent to a metric on $\H^5_\Z$ that is of negative type. We remark that~\cite{LN06} makes this assertion for a different metric on a larger continuous group that contains $\H_\Z^5$ as a discrete co-compact subgroup, but by a simple general result (e.g.~\cite[Theorem~8.3.19]{BBI01}) the word metric $d_W$ is bi-Lipschitz equivalent to the metric considered in~\cite{LN06}. Since $|\mathcal{B}_R|\asymp R^6$, we have $\sqrt{\log |\mathcal{B}_R|}\asymp\sqrt{\log R}$, so Theorem~\ref{thm:distortion R} implies Theorem~\ref{thm:main GL lower intro} through the duality result of Rabinovich that was recalled in Section~\ref{sec:embed intro}.

The following precise theorem about $L_1$ embeddings that need not be bi-Lipschitz implies Theorem~\ref{thm:distortion R} by considering the special case of the modulus $\upomega(t)=t/D$ for $D\ge 1$ and $t\in [0,\infty)$.
\begin{thm}\label{thm:integral criterion} There exists a universal constant $c\in (0,1)$ with the following property. Fix $R\ge 2$ and a nondecreasing function $\upomega:[1,\infty)\to [1,\infty)$. Then there exists $\upphi:\mathcal{B}_{R}\to L_1$ for which every distinct $x,y\in \mathcal{B}_{R}$ satisfy
\begin{equation}\label{eq:compression omega on ball}
\upomega\big(d_W(x,y)\big)\lesssim \|\upphi(x)-\upphi(y)\|_1\le d_W(x,y),
\end{equation}
{\bf if and only if}   $\omega(t)\lesssim t$ for all $t\in [1,\infty)$ and
\begin{equation}\label{eq:integral criterion}
\int_{1}^{c R} \frac{\upomega(s)^2}{s^3}\ud s\lesssim 1.
\end{equation}
\end{thm}
The fact that the integrability requirement~\eqref{eq:integral criterion} implies the existence of the desired embedding $\upphi$ is
 due to~\cite[Corollary~5]{Tes08}. The new content of Theorem~\ref{thm:integral criterion} is   that the existence of the embedding $\upphi$ implies~\eqref{eq:integral criterion}. By letting $R\to \infty$ in Theorem~\ref{thm:integral criterion} we see that there exists  $\upphi:\Z^5\to L_1$ that satisfies
 \begin{equation}\label{eq:compression assumption}
\forall\, x,y\in \Z^5,\quad  \upomega\big(d_W(x,y)\big)\lesssim \|\upphi(x)-\upphi(y)\|_1\le d_W(x,y),
\end{equation}
{\bf if and only if}
\begin{equation}\label{eq:1 to infty integral}
\int_{1}^\infty \frac{\upomega(s)^2}{s^3}\ud s\lesssim 1.
\end{equation}

In~\cite{CKN} it was shown that if $\upphi:\Z^5\to L_1$ satisfies~\eqref{eq:compression assumption},  then there must exist arbitrarily large $t\ge 2$ for which $\upomega(t)\lesssim t/(\log t)^\updelta$, where $\updelta>0$ is a universal constant. This follows from~\eqref{eq:1 to infty integral} with $\updelta=\frac12$, which is the largest possible constant for which this conclusion holds true.  This positively answers a question that was asked in~\cite[Remark~1.7]{CKN}.  In fact, it provides an even better conclusion, because~\eqref{eq:1 to infty integral} implies that, say, there must exist arbitrarily large $t\ge 4$ for which $$\upomega(t)\lesssim \frac{t}{\sqrt{(\log t)\log\log t}}.$$  (The precise criterion is the integrability condition~\eqref{eq:1 to infty integral}.) Finally, by considering $\upomega(t)=t^{1-\e}/D$ for $\e\in (0,1)$ and $D\ge 1$, we obtain the following notable corollary.

\begin{cor}[$L_1$ distortion of snowflakes]\label{coro:snoflake}
For every $\e\in (0,1)$ we have $c_1\big(\Z^5,d_W^{1-\e}\big)\asymp \frac{1}{\sqrt{\e}}$.
\end{cor}
The fact that for every $O(1)$-doubling metric space $(X,d)$ we have $c_1(X,d^{1-\e})\lesssim 1/\sqrt{\e}$ follows from an argument of~\cite{LMN05} (see also~\cite[Theorem~5.2]{NS11}). Corollary~\ref{coro:snoflake} shows that this is sharp. More generally, it follows from Theorem~\ref{thm:integral criterion} that for every $R\ge 2$ and $\e\in (0,1)$ we have
$$
c_1\big(\mathcal{B}_{R},d_W^{1-\e}\big)\asymp\min\left\{\frac{1}{\sqrt{\e}},\sqrt{\log R}\right\}.
$$

\subsection{Vertical-versus-horizontal isoperimetry}\label{sec:isoper intro}
Our new non-embeddability results are all consequences of an independently interesting isoperimetric-type inequality which we shall now describe.  Roughly speaking, this inequality subtly quantifies the fact that for any $n\in \Z$ and any $h\in \H_\Z^5$, there are many paths in the Cayley graph $\mathcal{X}_S(\H_\Z^{5})$ of length roughly $\sqrt{n}$ that connect $h$ to $h Z^n$.  Consequently, if a finite subset $\Omega\subset \Z^5$ has a small edge boundary in the Cayley graph, then the number of pairs $(x,y)\in \Z^5\times \Z^5$ for which $|\{x,y\}\cap \Omega|=1$ yet $x$ and $y$ differ only in their fifth (vertical) coordinate must also be small. It turns out that the proper interpretation of the term ``small'' is this context is not at all obvious, and it should be measured in a certain multi-scale fashion. Formally, we consider the following quantities.
\begin{defn}[Discrete boundaries]\label{def:discrete perimeters}
{ For $\Omega\subset \Z^{5}$, the {\bf horizontal boundary} of $\Omega$ is defined by
\begin{equation}\label{eq:def horizontal boundary}
\partial_{\hh}\Omega\eqdef \big\{(x,y)\in \Omega\times \left(\Z^{5}\setminus \Omega\right): x^{-1}y\in S\big\}.
\end{equation}
Given also  $t\in \N$, the $t${\bf-vertical boundary} of $\Omega$ is defined by
\begin{equation}\label{eq:def vertical t boundary}
\partial^t_{\vv} \Omega\eqdef \Big\{(x,y)\in \Omega\times \left(\Z^{5}\setminus \Omega\right): x^{-1}y\in \left\{Z^t,Z^{-t}\right\}\Big\}.
\end{equation}
The {\bf horizontal perimeter} of $\Omega$ is defined to be the cardinality $|\partial_{\hh}\Omega|$ of its horizontal boundary. The {\bf vertical perimeter} of $\Omega$ is defined to be the quantity
\begin{equation}\label{eq:def discrete vertical perimeter}
|\partial_{\vv}\Omega|\eqdef \bigg(\sum_{t=1}^\infty \frac{|\partial^t_{\vv}\Omega|^2}{t^2}\bigg)^{\frac12}.
\end{equation}
}
\end{defn}
The horizontal perimeter of $\Omega$ is nothing more than the size of its {\em edge boundary} in the Cayley graph $\mathcal{X}_S(\H_\Z^{5})$. The vertical perimeter of $\Omega$ is a more subtle concept that does not have such a simple combinatorial description. The definition~\eqref{eq:def discrete vertical perimeter} was first published in~\cite[Section~4]{LafforgueNaor}, where the isoperimetric-type conjecture that we resolve here as Theorem~\ref{thm:isoperimetric discrete}  below also appeared for the first time. These were formulated by the first named author and were circulating for several years before~\cite{LafforgueNaor} appeared, intended as a possible route towards the algorithmic application that we indeed succeed to obtain here. That ``vertical smallness'' should be measured through the quantity $|\partial_{\vv}\Omega|$, i.e.,  the $\ell_2$ norm of the sequence $\{|\partial_\vv^t\Omega|/t\}_{t=1}^\infty$, was arrived at through trial and error, inspired by  functional inequalities that were obtained in~\cite{AusNaoTes,LafforgueNaor}, as explained in~\cite[Section~4]{LafforgueNaor}.

\begin{thm}%[Vertical versus horizontal isoperimetric inequality]
\label{thm:isoperimetric discrete}  Every $\Omega\subset  \Z^{5}$ satisfies $|\partial_{\vv}\Omega|\lesssim |\partial_{\hh}\Omega|$.
\end{thm}
The significance of Theorem~\ref{thm:isoperimetric discrete} can only be  fully appreciated through an examination of the geometric and analytic reasons for its validity.  To facilitate this, we shall include in this extended abstract an extensive overview of the ideas of the proof of Theorem~\ref{thm:isoperimetric discrete}; see Section~\ref{sec:overview} below.  Before doing so, we shall now  demonstrate the utility of Theorem~\ref{thm:isoperimetric discrete}  by using it to deduce Theorem~\ref{thm:integral criterion}. As explained above, by doing so we shall conclude the proof of all of our new results (modulo Theorem~\ref{thm:isoperimetric discrete}), including the lower bound  on the integrality gap for the Goemans--Linial SDP.

\subsection{From isoperimetry to non-embeddability}  An equivalent formulation of  Theorem~\ref{thm:isoperimetric discrete} is  that every finitely supported function $\upphi:\Z^5\to L_1$ satisfies the following Poincar\'e-type inequality.
\begin{multline}\label{eq:discrete global intro}
\left(\sum_{t=1}^\infty \frac{1}{t^2}\bigg(\sum_{h\in \Z^5} \big\|\upphi\big(hZ^t\big)-\upphi(h)\big\|_1\bigg)^2\right)^{\frac12}\\\lesssim \sum_{h\in \Z^5}\sum_{\sigma\in S}\big\|\upphi(h\sigma)-\upphi(h)\big\|_1.
\end{multline}
Indeed, Theorem~\eqref{thm:isoperimetric discrete} is nothing more than the special case $\upphi=\1_\Omega$ of~\eqref{eq:discrete global intro}. Conversely, the fact that~\eqref{eq:discrete global intro}  follows from  Theorem~\eqref{thm:isoperimetric discrete} is a straightforward application of the cut-cone representation of $L_1$ metrics (see e.g.~\cite[Proposition~4.2.2]{DL97} or~\cite[Corollary~3.2]{Nao10}), though our proof will yield the (seemingly) stronger statement~\eqref{eq:discrete global intro} directly. Next, Section~3.2 of~\cite{LafforgueNaor} shows that~\eqref{eq:discrete global intro} formally implies its local counterpart, which asserts that there exists a universal constant $\alpha\ge 1$ such that for every $n\in \N$ and every $\upphi:\Z^5\to L_1$ we have
\begin{multline}\label{eq:discrete local intro}
\left(\sum_{t=1}^{n^2} \frac{1}{t^2}\bigg(\sum_{h\in \mathcal{B}_{n}} \big\|\upphi\big(hZ^t\big)-\upphi(h)\big\|_1\bigg)^2\right)^{\frac12}\\\lesssim \sum_{h\in \mathcal{B}_{\alpha n}}\sum_{\sigma\in S}\big\|\upphi(h\sigma)-\upphi(h)\big\|_1.
\end{multline}

To deduce Theorem~\ref{thm:integral criterion}, suppose that $R\ge 2$, that $\upomega:[0,\infty)\to [0,\infty)$ is nondecreasing and that the mapping $\upphi:\mathcal{B}_{R}\to L_1$ satisfies~\eqref{eq:compression omega on ball}. For notational convenience, fix two universal constants $\beta\in (0,1)$ and $\gamma\in (1,\infty)$ such that $\beta\sqrt{t}\le d_W(Z^t,\0)\le \gamma\sqrt{t}$ for every $t\in \N$. Note that~\eqref{eq:compression omega on ball} implies in particular that $\omega(R)\lesssim R$, so for every $c\in (0,1)$ the left hand side of~\eqref{eq:integral criterion} is at most a universal constant multiple of $R^2$.  Hence, it suffices to prove Theorem~\ref{thm:integral criterion} when $R\ge 1+\max\{\alpha,\gamma\}$, where $\alpha$ is the universal constant in~\eqref{eq:discrete local intro}. Denote $n=\lfloor \min\{R/(1+\gamma),(R-1)/\alpha\}\rfloor\in \N$. If $t\in \{1,\ldots,n^2\}$ and $h\in \mathcal{B}_n$ then $d_W(hZ^t,\0)\le n+\gamma \sqrt{t}\le (1+\gamma)n\le R$, and therefore we may apply~\eqref{eq:compression omega on ball} with $x=hZ^t$ and $y=h$ to deduce that $\|\upphi(hZ^t)-\upphi(h)\|_1\gtrsim \omega(d_W(Z^t,\0))\ge \omega(\beta\sqrt{t})$. Consequently,
\begin{align}\label{eq:pass to int}
&\sum_{t=1}^{n^2} \frac{1}{t^2}\bigg(\sum_{h\in \mathcal{B}_{n}} \big\|\upphi\big(hZ^t\big)-\upphi(h)\big\|_1\bigg)^2\gtrsim \sum_{t=1}^{n^2} \frac{|\mathcal{B}_n|^2\omega\big(\beta\sqrt{t}\big)^2}{t^2}\nonumber \\ \nonumber&\gtrsim n^{12}\sum_{t=1}^{n^2} \int_t^{t+1}\frac{\omega\big(\beta\sqrt{u/2}\big)^2}{u^2}\ud u=\beta^2n^{12}\int_{\frac{\beta}{\sqrt{2}}}^{\frac{\beta\sqrt{n^2+1}}{\sqrt{2}}}\frac{\omega(s)^2}{s^3}\ud s\\&\ge \frac{\beta^2(R/2)^{12}}{\max\{(1+\gamma)^{12},\alpha^{12}\}} \int_1^{\frac{\beta R }{2\max\{1+\gamma,\alpha\}}}\frac{\omega(s)^2}{s^3}\ud s,
\end{align}
where the second inequality in~\eqref{eq:pass to int} uses the fact that $\omega$ is non-decreasing, the penultimate step of~\eqref{eq:pass to int} uses the change of variable $s=\beta\sqrt{u/2}$, and for the final step of~\eqref{eq:pass to int} recall that $\beta<1$ and the definition of $n$.  At the same time, by our choice of $n$ we have $h\sigma\in \mathcal{B}_{\alpha n+1}\subset \mathcal{B}_R$ for every $h\in \mathcal{B}_{\alpha n}$ and $\sigma\in S$, and so by~\eqref{eq:compression omega on ball} we have $\|\upphi(h\sigma)-\upphi(h)\|_1\le d_W(h\sigma,h)=1$. The right hand side of~\eqref{eq:discrete local intro} is therefore at most a universal constant multiple of $|\mathcal{B}_{\alpha n}|\cdot|S|\lesssim (\alpha n)^6\lesssim R^6$. By contrasting~\eqref{eq:pass to int} with~\eqref{eq:discrete local intro} we obtain that the desired estimate~\eqref{eq:integral criterion} indeed holds true.

%\subsection{Previous work and directions for future research}\label{sec:previous} uniform demands

\subsection{Overview of the proof of Theorem~\ref{thm:isoperimetric discrete}}\label{sec:overview}

Our proof of~\eqref{eq:discrete global intro}, and hence also of Theorem~\ref{thm:isoperimetric discrete},  is carried out in a continuous setting that is equivalent to its discrete counterpart. Such a  passage from continuous to discrete is commonplace, and in the present setting this was carried out in~\cite{AusNaoTes,LafforgueNaor}. The  idea is to consider a continuous group that contains $\H_\Z^5$ and to deduce the discrete inequality~\eqref{eq:discrete global intro} from its (appropriately formulated) continuous counterpart via a partition of unity argument. There is an obvious way to embed $\H_\Z^5$ in a continuous group, namely by considering the same group of matrices as in~\eqref{eq:def H5}, but with the entries $\aaa,\bb,\cc,\dd,\ee$ now allowed to be arbitrary real numbers instead of integers. This is a indeed a viable route and the ensuing discussion could be carried out by considering the resulting continuous matrix group. Nevertheless, it is notationally  advantageous to work with a different (standard) realization of $\H_\Z^5$ which is isomorphic to the one that we considered thus far. We shall now introduce the relevant notation.

Fix  an orthonormal basis $\{X_1,X_2,Y_1,Y_2,Z\}$ of $\R^{5}$. If $h=\alpha_1X_1+\alpha_2X_2+\beta_1Y_1+\beta_2Y_2+\gamma Z\in \R^{5}$ then denote $x_i(h)=\alpha_i$, $y_i(h)=\beta_i$ for $i\in \{1,2\}$ and $z(h)=\gamma$, i.e., $x_1,x_2,y_1,y_2,z:\R^{5}\to \R$ are the coordinate functions corresponding to the above basis. The continuous Heisenberg group $\H^{5}$ is defined to be $\R^{5}$, equipped with the following group law.
 \begin{multline}\label{eq:def heisenberg algebra product}
 u v\eqdef u+v\\+\frac{x_1(u)y_1(v)-y_1(u)x_1(v)+x_2(u)y_2(v)-y_2(u)x_2(v)}{2} Z .
 \end{multline}

The identity element of $\H^{5}$ is $\0\in \R^{5}$ and the inverse of $h\in \R^{5}$ under the group law~\eqref{eq:def heisenberg algebra product} is equal to $-h$. By directly computing Jacobians, one checks that the Lebesgue measure on $\R^{5}$ is invariant under the group operation given in~\eqref{eq:def heisenberg algebra product}, i.e., it is a Haar measure of $\H^{5}$. In what follows, in order to avoid confusing multiplication by scalars with the group law of $\H^{5}$, for every $h\in \H^{5}$ and $t\in \R$ we shall use the exponential notation  $h^t=(th_1,\ldots,th_{5})$; this agrees with the group law when $t\in \Z$.  (This convention is not strictly necessary, but without it the ensuing discussion could  become somewhat notationally confusing.)

The subgroup of $\H^{5}$ that is generated by $\{X_1,X_2,Y_1,Y_2\}$ is the discrete Heisenberg group of dimension $5$, denoted $\H_\Z^{5}$. The apparent inconsistency with~\eqref{eq:def H5} is not an actual issue because it is straightforward to check that the two groups in question are in fact isomorphic. The linear span of $\{X_1,Y_1,Z\}$ is a subgroup of $\H^5$ which is denoted $\H^3$ (the $3$-dimensional Heisenberg group).

There is a canonical left-invariant metric on $\H^{5}$, commonly called the {\em Carnot--Carath\'eodory metric}, which we denote  by $d$. We refer to \cite{CDPT07}  for a precise definition of this metric. For the purpose of the present discussion it suffices to know that $d$ possesses the following properties.  Firstly, for every $g,h\in \H^{5}$ and $\uptheta\in \R$ we have $d(\mathfrak{s}_\uptheta (g),\mathfrak{s}_\uptheta (h))=|\uptheta| d(g,h)$. Here, $\mathfrak{s}_\uptheta$ denotes the {\em Heisenberg scaling} by $\uptheta$, given by the formula
$$\mathfrak{s}_\uptheta (\alpha_1,\alpha_2,\beta_1,\beta_2,\gamma)=(\uptheta\alpha_1,\uptheta \alpha_2,\uptheta \beta_1,\uptheta \beta_2,\uptheta^2\gamma)$$
% \begin{multline*}
% \mathfrak{s}_\uptheta (\alpha_1X_1+\alpha_2X_2+\beta_1Y_1+\beta_2Y_2+\gamma Z)\\\eqdef \uptheta (\alpha_1X_1+\alpha_2X_2+\beta_1Y_1+\beta_2Y_2)+\uptheta^2\gamma Z
%  \end{multline*}
Secondly, the restriction of $d$ to the subgroup $\H_\Z^{5}$ is bi-Lipschitz to the word metric induced by its generating set $\{X_1^{\pm 1}, X_2^{\pm 1},Y_1^{\pm 1},Y_2^{\pm 1}\}$. Thirdly,  there exists $C\in (1,\infty)$ such that  every $h\in \H^5$ satisfies
\begin{equation}\label{eq:cc normalization}
 d(h,\0)\le |x_1(h)|+|x_2(h)|+|y_1(h)|+|y_2(h)|+4\sqrt{|z(h)|}\le \frac{C}{2}d(h,\0).
\end{equation}

Given $r\in (0,\infty)$ we shall denote by $B_r\subset \H^{5}$ the open ball in the metric $d$ of radius $r$ centered at the identity element, i.e., $B_r=\{h\in \H^{5}:\ d(\0,h)< r\}$. For $\Omega\subset \H^5$ the Lipschitz constant of a mapping $f:\Omega\to \R$ relative to the metric $d$ will be denoted by $\|f\|_{\Lip(\Omega)}$. For  $s\in (0,\infty)$, the notation $\CH^s$ will be used exclusively to denote the $s$-dimensional Hausdorff measure that is induced by the metric $d$ (see e.g.~\cite{Mat95}). One checks that $\CH^{6}$ is proportional to the Lebesgue measure on $\R^{5}$ and that the restriction of $\CH^4$ to the subgroup $\H^3$ is proportional to the Lebesgue measure on $\H^3$ (under the canonical identification of  $\H^3$ with $\R^3$). For two measurable subsets $E,U\subset \H^{5}$ define the {\em normalized vertical perimeter} of $E$ in $U$ to be the function $\overline{\mathsf{v}}_U(E):\R\to [0,\infty]$ given by setting for every $s\in \R$,
\begin{align}\label{eq:def normalized vertical}
\nonumber\overline{\mathsf{v}}_U(E)(s)&\eqdef \frac{1}{2^{s}}\CH^{6}\Big( \big(E\symdiff \big(E Z^{2^{2s}}\big)\big)\cap U\Big)\\&=\frac{1}{2^{s}}\int_{U} \Big|1_E(u)-1_{E}\big(uZ^{-2^{2s}}\big)\Big|\ud \CH^6(u).
\end{align}
where $A\symdiff B\eqdef (A\setminus B)\cup (B\setminus A)$ is the symmetric difference.  We also denote $\overline{\mathsf{v}}(E)\eqdef\overline{\mathsf{v}}_{\H^5}(E)$.

%One checks that $\CH^{6}$ is proportional to the Lebesgue measure on $\R^{5}$ and one has $\CH^{6}(B_r)=r^{6}\CH^{6}(B_1)$ for every $r\in (0,\infty)$. Also, one can check that the %restriction of $\CH^4$ to the subgroup $\H^3$ is proportional to the Lebesgue measure on $\H^3$ when one identified $\H^3$ with $\R^3$.

The isoperimetric-type inequality of Theorem~\ref{thm:v vs h cont intro} below implies Theorem~\ref{thm:isoperimetric discrete}. See \cite{NY-full} for an explanation of this (standard) deduction; the argument is a straightforward use of the co-area formula (see e.g.~\cite{Amb01,Mag11}) to pass from sets to functions, followed by the partition of unity argument of~\cite[Section~3.3]{LafforgueNaor} to pass from the continuous setting to the desired discrete inequality~\eqref{eq:discrete global intro}.
 \begin{thm}\label{thm:v vs h cont intro} $
\big\|\overline{\mathsf{v}}(E)\big\|_{L_2(\R)}\lesssim \CH^{5}(\partial E)
$
for all open $E\subset \H^{5}$.
\end{thm}

We shall now explain the overall strategy and main ideas of our proof of Theorem~\ref{thm:v vs h cont intro}.  Complete technical details are included in \cite{NY-full}. A key new ingredient appears in Section~\ref{sec:into lip graph} below, which is the {\em only} place in our proof where we use the fact that we are dealing with $\H^5$ rather than $\H^3$.  In fact, the analogue of Theorem~\ref{thm:v vs h cont intro} for $\H^3$ (i.e., with $\CH^5(\partial E)$ replaced by $\CH^3(\partial E)$ and $\overline{\mathsf{v}}(E)(\cdot)$ defined in the same way as in~\eqref{eq:def normalized vertical} but with $\CH^6$ replaced by the restriction of $\CH^4$ to $\H^3$) is {\em false} (see Section~\ref{sec:3d} below). The crux of the matter is the special case of Theorem~\ref{thm:v vs h cont intro} where the boundary of $E$ is (a piece of) an {\em intrinsic Lipschitz graph}. Such sets were introduced by Franchi, Serapioni, and Serra Cassano~\cite{FSSC06}.  These sets can be quite complicated, and in particular they {\em are not the same} as graphs of functions (in the usual sense)  that are Lipschitz with respect to the Carnot--Carath\'eodory metric. Our proof of this special case relies crucially on an $L_2$-variant of~\eqref{eq:discrete global intro} for $\H^3$ that was proven in~\cite{AusNaoTes} using representation theory and in~\cite{LafforgueNaor} using Littlewood--Paley theory. In essence, our argument ``lifts'' a certain $L_2$ inequality in lower dimensions to a formally stronger endpoint $L_1$ (or isoperimetric-type) inequality in higher dimensions. Once the special case is established, we prove Theorem~\ref{thm:v vs h cont intro} in its full generality by decomposing an open set $E$ into parts whose boundaries are close to pieces of intrinsic Lipschitz graphs and applying the special case to each part of this decomposition.  We deduce the desired estimate by summing up all the inequalities thus obtained. Such a ``corona decomposition'' is an important and widely-used tool in harmonic analysis on $\R^n$ that was formulated by David and Semmes in~\cite{DavidSemmesSingular}. For the present purpose we need to devise an ``intrinsic version'' of a corona decomposition on the Heisenberg group. This step uses a different ``coercive quantity''  to control local overlaps, but for the most part it follows  the lines of the well-understood methodology of David and Semmes, as described in the monographs~\cite{DavidSemmesSingular,DSAnalysis}.

\subsubsection{Intrinsic Lipschitz graphs}\label{sec:into lip graph} Set $V\eqdef\{h\in \H^5:\ x_2(h)=0\}$. For $f:V\to \R$ define
\begin{multline}\label{eq:lip graph abcd}
\Gamma_f\eqdef \Big\{vX_2^{f(v)}:\ v\in V\Big\}\\\stackrel{\eqref{eq:def heisenberg algebra product}}{=} \Big\{\Big(\aaa,f(\aaa,\cc,\dd,\ee),\cc,\dd,\ee-\frac12 \dd f(\aaa,\cc,\dd,\ee) \Big):\ \aaa,\cc,\dd,\ee\in \R\Big\},
\end{multline}
where~\eqref{eq:lip graph abcd}  uses the identification of $\aaa X_1+\bb X_2+\cc Y_1+\dd Y_2+\ee Z\in \H^5$ with $(\aaa,\bb,\cc,\dd,\ee)\in \R^5$ and the identification of  $(\aaa,0,\cc,\dd,\ee)\in V$ with $(\aaa,\cc,\dd,\ee)\in \R^4$ (thus we think of the domain of $f$ as equal to $\R^4$). The set $\Gamma_f$ is a typical  {\em intrinsic graph} in $\H^5$.  See \cite{NY-full} for a discussion of the general case, which is equivalent to this case via a symmetry of $\H^5$ (so the ensuing discussion has no loss of generality). Suppose that $\lambda\in (0,\infty)$. We say that $\Gamma_f$ is an {\em intrinsic $\lambda$-Lipschitz graph} over the vertical hyperplane $V$ if
\begin{equation}\label{eq:lambda lip condition}
\forall\, w_1,w_2\in \Gamma_f,\qquad |x_2(w_1)-x_2(w_2)|\le \lambda d(w_1,w_2).
 \end{equation}
 Due to~\eqref{eq:lip graph abcd} the condition~\eqref{eq:lambda lip condition} amounts to a point-wise inequality for $f$ that is somewhat complicated, and in particular it does not imply that $f$ must be Lipschitz with respect to the restriction of the Carnot--Carath\'eodory metric to the hyperplane $V$, as explained in~\cite[Remark~3.13]{FSSCDifferentiability}.

Denote by $\Gamma_f^+=\{vX_2^t:\ v\in V\ \wedge \  t>f(v)\}$ the half-space that is bounded by the intrinsic graph $\Gamma_f$. Suppose that  $\Gamma_f$ is an intrinsic $\lambda$-Lipschitz graph with $\lambda\in (0,1)$.  We claim that
\begin{equation}\label{eq:lip graph ineq intro}
\forall\, r\in (0,\infty),\qquad \big\|\overline{\mathsf{v}}_{B_r}\big(\Gamma^+_f\big)\big\|_{L_2(\R)}\lesssim \frac{r^5}{1-\lambda}.
\end{equation}
When, say, $\lambda\in (0,\frac12)$, the estimate~\eqref{eq:lip graph ineq intro} is in essence the special case of Theorem~\ref{thm:v vs h cont intro} for pieces of Lipschitz graphs. This is so because, due to the isoperimetric inequality for the Heisenberg group~\cite{Pan82}, the right-hand side of~\eqref{eq:lip graph ineq intro}  is at most a universal constant multiple of $\CH^5(\partial(B_r\cap \Gamma_f^+))$ whenever $\CH^6(B_r\cap \Gamma_f^+)\gtrsim r^6$, i.e., provided that $\Gamma_f^+$ occupies a constant fraction of the volume of the ball $B_r$. The estimate~\eqref{eq:lip graph ineq intro} will be used below only in such a non-degenerate situation.

 The advantage of working in $\H^5$ rather than $\H^3$  is that $V\subset \H^5$ can be sliced into copies of $\H^3$.  We will bound $\|\overline{\mathsf{v}}_{B_r}(\Gamma^+_f)\|_{L_2(\R)}$ by decomposing $\Gamma_f^+$ into a corresponding family of slices. Write
\begin{equation}\label{eq:def h(u)}
\forall\, u\in \H^5,\quad h_u\eqdef X_1^{x_1(u)}+Y_1^{y_1(u)}+Z^{z(u)+\frac12 x_2(u)y_2(u)}\in \H^3.
\end{equation}
Recalling~\eqref{eq:def heisenberg algebra product}, one computes directly that $u=Y_2^{y_2(u)}h_uX_2^{x_2(u)}$. Let $C\in (1,\infty)$ be the universal constant in~\eqref{eq:cc normalization}. A straightforward computation using~\eqref{eq:cc normalization} shows that $d(h_u,\0)\le Cd(u,\0)$. Also, \eqref{eq:cc normalization} implies that $|y_2(u)|\le Cd(u,\0)$. These simple observations demonstrate that
\begin{equation}\label{eq:ball product set}
\forall\, u\in \H^5,\qquad \1_{B_r}(u)\le \1_{[-Cr,Cr]}\big(y_2(u)\big)\1_{\H^3\cap B_{Cr}}(h_u).
\end{equation}

For every $\chi\in \R$ define $f_\chi:\H^3\to \R$ by $f_\chi(h)=f(Y_2^\chi h)$ (recall that $\H^3$ is the span of $\{X_1,Y_1,Z\}$, so $Y_2^\chi h\in V$ is in the domain of $f$). Under this notation $u\in \Gamma^+_f$ if and only if $x_2(u)>f_{y_2(u)}(h_u)$. Also, for every $\alpha\in \R$ we have $uZ^\alpha\in \Gamma^+_f$ if and only if $x_2(u)>f_{y_2(u)}(h_u Z^\alpha)$, since $h_{uZ^{\alpha}}=h_u Z^\alpha$ by~\eqref{eq:def h(u)}.  Due to~\eqref{eq:def normalized vertical} and~\eqref{eq:ball product set}, these observations imply that for every $s\in \R$ we have
\begin{align}\label{eq:estimate on product instead of ball}
\nonumber &\overline{\mathsf{v}}_{B_r}\big(\Gamma^+_f\big)(s)
\\&\nonumber \le \frac{1}{2^s}\int_{\H^5} \Big|\1_{\{x_2(u)>f_{y_2(u)}(h_u)\}}-\1_{\{x_2(u)>f_{y_2(u)}(h_uZ^{-2^{2s}})\}}\Big|\nonumber\\&\qquad\qquad \times\1_{[-Cr,Cr]}\big(y_2(u)\big)\1_{\H^3\cap B_{Cr}}(h_u)\ud \CH^6(u).
\end{align}
Recall that $\CH^6$ is proportional to the Lebesgue measure on $\H^5$. Hence, if we continue to canonically identify $\aaa X_1+\bb X_2+\cc Y_1+\dd Y_2+\ee Z\in \H^5$ with $(\aaa,\bb,\cc,\dd,\ee)\in \R^5$ and $\aaa X_1+\cc Y_1+\ee Z\in \H^3$ with $(\aaa,\cc,\ee)\in \R^3$ then, recalling~\eqref{eq:def h(u)}, the integral in the right hand side of~\eqref{eq:estimate on product instead of ball} is proportional to
\begin{align*}
&\int_{\R^5} \Big|\1_{\big\{\bb>f_{\dd}\big(\aaa,\cc,\ee+\frac12\bb \dd\big)\big\}}-\1_{\big\{\bb>f_{\dd}\big(\aaa,\cc,\ee+\frac12\bb \dd-2^{2s}\big)\big\}}\Big| \\&\qquad\qquad\times \1_{[-Cr,Cr]}(\dd)\1_{\H^3\cap B_{Cr}}\Big(\aaa,\cc,\ee+\frac12\bb \dd\Big)\ud(\aaa,\bb,\cc,\dd,\ee)\\
&= \int_{\R^5} \Big|\1_{\{\bb>f_{\dd}(\alpha,\gamma,\upepsilon)\}}-\1_{\{\bb>f_{\dd}(\alpha,\gamma,\upepsilon-2^{2s})\}}\Big|\\&\qquad\qquad\times \1_{[-Cr,Cr]}(\dd)\1_{\H^3\cap B_{Cr}}(\alpha,\gamma,\upepsilon)\ud(\alpha,\bb,\gamma,\dd,\upepsilon),
\end{align*}
 where for each fixed $\bb,\dd\in \R$ we made the change of variable $(\alpha,\gamma,\upepsilon)=(\aaa,\cc,\ee+\bb \dd/2)$. Since the restriction of the Hausdorff measure $\CH^4$ to $\H^3$ is proportional to the Lebesgue measure on $\H^3\cong \R^3$, we conclude from the above considerations that for every $s\in \R$ we have
\begin{align}\label{eq:v bar bound at s}
\nonumber &\overline{\mathsf{v}}_{B_r}\big(\Gamma^+_f\big)(s)\\\nonumber &\lesssim \frac{1}{2^s} \int_{-Cr}^{Cr}\int_{\H^3\cap B_{Cr}}\\&\qquad \bigg(\int_{-\infty}^\infty \Big|\1_{\{\xi>f_{\chi}(h)\}}-\1_{\{\xi>f_{\chi}(hZ^{-2^{2s}})\}}\Big|\ud \xi\bigg)\ud \CH^4(h)\ud \chi\nonumber\\
&= \frac{1}{2^s} \int_{-Cr}^{Cr}\int_{\H^3\cap B_{Cr}} \Big|f_\chi(h)-f_\chi\big(hZ^{-2^{2s}}\big)\Big|\ud \CH^4(h)\ud \chi.
\end{align}

Next, fix  $h_1,h_2\in \H^3$ and $\chi\in \R$. Denote $w_1\eqdef Y_2^\chi h_1X_2^{f_\chi(h_1)}$ and $w_2\eqdef Y_2^\chi h_2X_2^{f_\chi(h_2)}$. By design we have $w_1,w_2\in \Gamma_f$ and therefore  we may   apply~\eqref{eq:lip graph ineq intro} to deduce that
\begin{align}\label{eq:to deduce lip sections}
\nonumber &|f_\chi(h_1)-f_\chi(h_2)|=|x_2(w_1)-x_2(w_2)|\le \lambda d(w_1,w_2)\\ \nonumber&=\lambda d\Big(Y_2^\chi h_1X_2^{f_\chi(h_1)},Y_2^\chi h_2X_2^{f_\chi (h_2)}\Big)
\\ \nonumber&=\lambda d\Big(\0,h_1^{-1}h_2X_2^{f_\chi(h_2)-f_\chi(h_1)}\Big)\nonumber \\&\le \lambda\Big(Cd(h_1,h_2)+|f_\chi(h_1)-f_\chi(h_2)|\Big),
\end{align}
where the first inequality in~\eqref{eq:to deduce lip sections} uses~\eqref{eq:lip graph ineq intro} , the penultimate step of~\eqref{eq:to deduce lip sections} uses the left-invariance of the metric $d$ and the fact that $X_2$ commutes with all of the elements of $\H^3$, and the final step of~\eqref{eq:to deduce lip sections} uses~\eqref{eq:cc normalization} (twice).  The estimate~\eqref{eq:to deduce lip sections} simplifies to show that $|f_\chi(h_1)-f_\chi(h_2)|\lesssim d(h_1,h_2)/(1-\lambda)$, i.e., for every fixed $\chi\in \R$ the function $f_\chi$ is Lipschitz on $\H^3$ with  $\|f_\chi\|_{\Lip(\H^3)}\lesssim 1/(1-\lambda)$.

In~\cite[Theorem~7.5]{AusNaoTes} the following inequality was proved for a Lipschitz function $\psi:\H^3\to \R$ and $\rho\in (0,\infty)$ as a consequence of a continuous $L_2$-variant of~\eqref{eq:discrete global intro}.  Due to its quadratic nature, this variant can be proved using a decomposition into irreducible representations (i.e., a spectral argument).

%A  generalization of this result (an $L_p$ version for every $p\in (1,\infty)$  as well as more general target spaces) follows from~\cite[Theorem~2.1]{LafforgueNaor}, where it is proven %using Littlewood--Paley theory.
\begin{equation}\label{eq:psi ineq}
\int_0^{\rho^2}\int_{B_\rho\cap \H^3} \left|\psi(h)-\psi\big(hZ^{-t}\big)\right|^2\ud\mathcal{H}^4(h)\frac{\ud t}{t^2}\lesssim \rho^4\|\psi\|_{\Lip(\H^3)}^2.
\end{equation}
Consequently,
%Fix $s\in \R$ and apply~\eqref{eq:psi ineq} with $\psi=f_s$ while using the above bound on $\|f_s\|_{\Lip(\H^3)}$. This gives
\begin{align}
&\label{eq:use slice lip}\frac{r^5}{(1-\lambda)^2}\\ \nonumber&\gtrsim  \int_{-Cr}^{Cr}\int_0^{(Cr)^2}\int_{B_{Cr}\cap \H^3} \left|f_\chi(h)-f_\chi\big(hZ^{-t}\big)\right|^2\ud\mathcal{H}^4(h)\frac{\ud t}{t^2}\ud \chi \\&=
\int_{-\infty}^{\log_2(Cr)}\frac{2\log 2}{2^{2s}}\nonumber\\&\quad \times\int_{-Cr}^{Cr}\int_{B_{Cr}\cap \H^3} \left|f_\chi(h)-f_\chi\big(hZ^{-2^{2s}}\big)\right|^2\ud\CH^4(h)\ud \chi\ud s \label{eq:change of variable}\\
&\label{eq:use CS on slice} \gtrsim \int_{-\infty}^{\log_2(Cr)} \frac{1}{r^5}\\ \nonumber &\quad\times \bigg(\frac{1}{2^s}\int_{-Cr}^{Cr}\int_{B_{Cr}\cap \H^3} \left|f_\chi(h)-f_\chi\big(hZ^{-2^{2s}}\big)\right|\ud\CH^4(h)\ud \chi\bigg)^2 \ud s \\
&\gtrsim \frac{1}{r^5} \int_{-\infty}^{\log_2(Cr)}\overline{\mathsf{v}}_{B_r}\big(\Gamma^+_f\big)(s)^2 \ud s,\label{eq:use bound on v bar}
\end{align}
In~\eqref{eq:use slice lip} we applied~\eqref{eq:psi ineq} with $\psi=f_\chi$ for each $\chi\in [-Cr,Cr]$, while using $\|f_\chi\|_{\Lip(\H^3)}\lesssim 1/(1-\lambda)$. In~\eqref{eq:change of variable} we made the change of variable $t=2^{2s}$. In~\eqref{eq:use CS on slice} we used the Cauchy--Schwarz inequality while noting that $\CH^4(B_{Cr}\cap \H^3)\asymp r^4$. Finally, \eqref{eq:use bound on v bar} follows from an application of~\eqref{eq:v bar bound at s}. Now,
\begin{align}\label{eq:got desired graph}
&\nonumber \big\|\overline{\mathsf{v}}_{B_r}\big(\Gamma^+_f\big)\big\|_{L_2(\R)}^2\\ \nonumber&=\int_{-\infty}^{\log_2(Cr)}\overline{\mathsf{v}}_{B_r}\big(\Gamma^+_f\big)(s)^2 \ud s+\int_{\log_2(Cr)}^\infty\overline{\mathsf{v}}_{B_r}\big(\Gamma^+_f\big)(s)^2 \ud s\\
&\lesssim \frac{r^{10}}{(1-\lambda)^2}+\int_{\log_2(Cr)}^\infty\frac{\CH^6(B_r)^2}{2^{2s}} \ud s \nonumber\\&\asymp\frac{r^{10}}{(1-\lambda)^2}+\int_{\log_2(Cr)}^\infty \frac{r^{12}}{2^{2s}}\ud s\asymp \frac{r^{10}}{(1-\lambda)^2},
\end{align}
where we estimated the second integral using the trivial bound $\overline{\mathsf{v}}_{B_r}(E)(s)\le \CH^6(B_r)/2^s\asymp r^6/2^s$. By taking square roots of both sides of~\eqref{eq:got desired graph} we obtain the desired estimate~\eqref{eq:lip graph ineq intro}. It is important to stress that this proof does not work for functions on $\H^3$ because it relies on slicing $\H^5$ into copies of $\H^3$.  There is no analogue of~\eqref{eq:psi ineq} for $1$-dimensional vertical slices of $\H^3$.

\subsubsection{An intrinsic corona decomposition} In Section~\ref{sec:into lip graph} we presented the complete details of the proof of a crucial new ingredient that underlies the validity of Theorem~\ref{thm:v vs h cont intro}.  This ingredient is the {\em only step} that relies on a property of $\H^5$ that is not shared by $\H^3$. We believe that it is important to fully explain this key ingredient within this extended abstract, but this means that we must defer the details of the formal derivation of Theorem~\ref{thm:v vs h cont intro} from its special case~\eqref{eq:lip graph ineq intro} to the full version~\cite{NY-full}.  The complete derivation requires additional terminology and notation, but the main idea is to produce ``intrinsic corona decompositions'' in the Heisenberg group.  Corona decompositions are an established tool in analysis for reducing the study of certain singular integrals on $\R^n$ to the case of Lipschitz graphs, starting with seminal works of David~\cite{Dav84,Dav91} and Jones~\cite{Jon89,Jon90} on the Cauchy integral and culminating with the David--Semmes theory of quantitative rectifiability~\cite{DavidSemmesSingular,DSAnalysis}.  Our adaptation of this technique is mostly technical, but it will also involve a conceptually new ingredient, namely the use of quantitative monotonicity for this purpose. We will now outline the remainder of the proof of Theorem~\ref{thm:v vs h cont intro}.

Our arguments hold for Heisenberg groups of any dimension (including $\H^3$), but we avoid introducing new notation by continuing to work with $\H^5$ for now. The first step is to show that in order to establish Theorem~\ref{thm:v vs h cont intro} it suffices to prove that for every $r\in (0,\infty)$ and every $E\subset \H^5$, we have $\|\overline{\mathsf{v}}_{B_r}(E)\|_{L_2(\R)}\lesssim r^5$ under the additional assumption  that the sets $E$, $\H^5\setminus E$, and $\partial E$ are $r$-locally Ahlfors-regular, i.e.,  $\CH^6(uB_\rho \cap E)\asymp \rho^6\asymp \CH^6(vB_\rho \setminus E)$ and $\CH^5(wB_\rho\cap \partial E)\asymp \rho^5$  for all $\rho\in (0,r)$ and $(u,v,w)\in E\times (\H^5\setminus E)\times \partial E$.  %This step is mostly technical, relying only on the fact that the Heisenberg group is a doubling metric space that satisfies a Poincaré inequality.
We prove this  by first applying a Heisenberg scaling and an approximation argument to reduce Theorem~\ref{thm:v vs h cont intro} to the case that $E$ is a ``cellular set,'' i.e., it is a union of parallelepipeds of the form $h[-\frac12,\frac12]^5$ as $h$ ranges over a subset of the discrete Heisenberg group $\H_\Z^5\subset \H^5$.  Any such set is Ahlfors-regular on sufficiently small balls.  We next argue that $E$ can be decomposed into sets that satisfy the desired local Ahlfors-regularity.  The full construction of this decomposition appears in~\cite{NY-full}, but we remark briefly that it amounts to the following natural ``greedy'' iterative procedure. If one of the sets $E, \H^5\setminus E, \partial E$ were not locally Ahlfors-regular then there would be some smallest ball $B$ such that the density of $E$, $\H^5\setminus E$ or $\partial E$ is either too low or too high on $B$. By replacing $E$ by either $E\cup B$ or $E\setminus B$, we cut off a piece of $\partial E$ and decrease $\CH^5(\partial E)$.  Since $B$ was the smallest ball where Ahlfors-regularity fails, $E, \H^5\setminus E, \partial E$ are Ahlfors-regular on balls smaller than $B$.  Repeating this process eventually reduces $E$ to the empty set, and we arrive at the conclusion of Theorem~\ref{thm:v vs h cont intro} for the initial set $E$ by proving the (local version of) the theorem for each piece of this decomposition, then summing the resulting inequalities.  We will therefore suppose from now on that $E$, $\H^5\setminus E$ and $\partial E$ are all locally Ahlfors-regular.

The next step is the heart of the matter: approximating $\partial E$ by intrinsic Lipschitz graphs so that we can use the fact that Theorem~\ref{thm:v vs h cont intro} holds for (pieces of) such graphs.  The natural way to do this is to construct (an appropriate Heisenberg version of) a corona decomposition in the sense of~\cite{DavidSemmesSingular,DSAnalysis}.  Such a decomposition covers $\partial E$ by two types of sets, called \emph{stopping-time regions} and \emph{bad cubes}.  Stopping-time regions correspond to parts of $\partial E$ that are close to intrinsic Lipschitz graphs, and bad cubes correspond to parts of $\partial E$, like sharp corners, that are not.  The multiplicity of this cover depends on the shape of $\partial E$ at different scales.  For example, $\partial E$ might look smooth on a large neighborhood of a point $x$, jagged at a medium scale, then smooth again at a small scale.  If so, then $x$ is contained in a large stopping-time region, a medium-sized bad cube, and a second small stopping-time region.  A cover like this is a corona decomposition if it satisfies a {\em Carleson packing condition} (see~\cite{NY-full}) that bounds its average multiplicity on any ball.

We construct our cover following the well-established methods of~\cite{DavidSemmesSingular,DSAnalysis}.  We start by constructing a sequence of nested partitions of $\partial E$ into pieces called \emph{cubes}; this is a standard construction due to Christ~\cite{ChristTb} and David~\cite{DavidWavelets} and only uses the Ahlfors regularity of $\partial E$.  These partitions are analogues of the standard tilings of $\R^n$ into dyadic cubes.  Next, we classify the cubes into \emph{good cubes}, which are close to a piece of a hyperplane, and \emph{bad cubes}, which are not.  In order to produce a corona decomposition, there cannot be too many bad cubes, i.e., they must satisfy a Carleson packing condition.  In~\cite{DavidSemmesSingular,DSAnalysis}, this condition follows from \emph{quantitative rectifiability}; the surface in question is assumed to satisfy a condition that bounds the sum of its (appropriately normalized) local deviations from hyperplanes.  These local deviations are higher-dimensional versions of Jones' $\beta$-numbers~\cite{Jon89,Jon90}, and the quantitative rectifiability assumption leads to the desired packing condition.  In the present setting, the packing condition follows instead from {\em quantitative non-monotonicity}.  The concept of the quantitative non-monotonicity of a set $E\subset \H^5$ (see~\cite{NY-full}) was first defined in~\cite{CKN09,CKN},  where the kinematic formula for the Heisenberg group was used to show that the total non-monotonicity of all of the cubes is at most a constant multiple of $\CH^5(\partial E)$.  This means that there cannot be many cubes that have large non-monotonicity. By a result of~\cite{CKN09,CKN}, if a set has small non-monotonicity, then its boundary is close to a hyperplane.  Consequently, most cubes are close to hyperplanes and are therefore good.  (The result in~\cite{CKN09,CKN} is stronger than what we need for this proof; it provides power-type bounds on how closely a nearly-monotone surface approximates a hyperplane.  For our purposes, it is enough to have \emph{some} bound (not necessarily power-type) on the shape of nearly-monotone surfaces, and we can deduce the bound that we need by applying a quick compactness argument to a result from~\cite{CheegerKleinerMetricDiff} that states that if a set is {\em precisely monotone} (i.e., every line intersects its boundary in at most one point), then it is a half-space.)

Next, we partition the good cubes into stopping-time regions by using an iterative construction that corrects overpartitioning that may have occurred when the Christ cubes were constructed.  If $Q$ is a largest good cube that hasn't been treated yet and if $P$ is its approximating half-space, we find all of the descendants of $Q$ with approximating half-spaces that are sufficiently close to $P$.  If we glue these half-spaces together using a partition of unity, the result is an intrinsic Lipschitz half-space that approximates all of these descendants.  By repeating this procedure for each untreated cube, we obtain a collection of stopping-time regions.  These regions satisfy a Carleson packing condition because if a point $x\in \partial E$ is contained in many different stopping-time regions, then either $x$ is contained in many different bad cubes, or $x$ is contained in good cubes whose approximating hyperplanes point in many different directions.  In either case, these cubes generate  non-monotonicity, so there can only be a few points with large multiplicity.

The construction above leads to the proof of Theorem~\ref{thm:v vs h cont intro} as follows.  The vertical perimeter of $\partial E$ comes from three sources: the bad cubes, the approximating Lipschitz graphs, and the error incurred by approximating a stopping-time region by an intrinsic Lipschitz graph.  By the Carleson packing condition, there are few bad cubes, and they contribute vertical perimeter on the order of $\cH^5(\partial E)$.  By the result of Section~\ref{sec:into lip graph}, the intrinsic Lipschitz graphs also contribute vertical perimeter on the order of $\cH^5(\partial E)$.  Finally, the vertical perimeter of the difference between a stopping-time region and an intrinsic Lipschitz graph is bounded by the size of the stopping-time region.  The stopping-time regions also satisfy a Carleson packing condition, so these errors also contribute vertical perimeter on the order of $\cH^5(\partial E)$.  Summing these contributions, we obtain the desired bound.

\subsection{Historical overview and directions for further research}\label{sec:previous}  Among the well-established deep and multifaceted connections between theoretical computer science and pure mathematics, the Sparsest Cut Problem stands out for its profound and often unexpected impact on a variety of areas. Indeed, previous research on  this question came hand-in-hand with the development of remarkable mathematical and algorithmic ideas that spurred many further works of importance in their own right. Because the present work belongs to this tradition, we will try to put it into context by elaborating further on the history of these investigations and describing directions for further research and open problems.  Some of these directions will appear in forthcoming work.

The first polynomial-time algorithm for Sparsest Cut with approximation ratio  $O(\log n)$ was obtained in the important work~\cite{LR99}, which studied the notable special case of {\em Sparsest Cut with Uniform Demands} (see Section~\ref{sec:uniform}  below).  This work introduced a linear programming relaxation and developed influential techniques for its analysis, and it has led to a myriad of algorithmic applications. The seminal contributions~\cite{LLR95,AR98} obtained the upper bound $\rho_{\mathsf{GL}}(n)\lesssim \log n$ in full generality by incorporating a classical embedding theorem of Bourgain~\cite{Bou85}, thus heralding the transformative use of metric embeddings in algorithm design. The matching lower bound on the integrality gap of this linear program was proven in~\cite{LR99,LLR95}.  This showed for the first time that Bourgain's embedding theorem is asymptotically sharp and was the first demonstration of the power of expander graphs in the study of metric embeddings.

A $O(\sqrt{\log n})$ upper bound for the approximation ratio of the Goemans--Linial algorithm in the  case of uniform demands was obtained in the important work~\cite{ARV09}.  This work relied on a clever use of the concentration of measure phenomenon and introduced influential techniques such as a ``chaining argument'' for metrics of negative type and the use of expander flows. \cite{ARV09} also had direct impact on results in pure mathematics, including combinatorics and metric geometry; see e.g.~the ``edge replacement theorem'' and the estimates on the observable diameter of doubling metric measure spaces in~\cite{NRS05}. The best-known upper bound $\rho_{\mathsf{GL}}(n)\lesssim (\log n)^{\frac12+o(1)}$ of~\cite{ALN08} built on the (then very recent) development of two techniques: The chaining argument of~\cite{ARV09} (through its refined analysis in~\cite{Lee05}) and the {\em measured descent} embedding method of~\cite{KLMN05} (through its statement as a gluing technique for Lipschitz maps in~\cite{Lee05}). Another important input to~\cite{ALN08} was a re-weighting argument of~\cite{CGR08} that allowed for the construction of an appropriate ``random zero set'' from the argument of~\cite{ARV09,Lee05} (see~\cite{Nao10,Nao14} for more on this notion and its significance).

The impossibility result~\cite{KV15} that refuted the Goemans--Linial conjecture relied on a striking link to complexity theory through the Unique Games Conjecture (UGC), as well as an interesting use of discrete harmonic analysis (through~\cite{Bou02}) in this context; see also~\cite{KR09} for an incorporation of a different tool from discrete harmonic analysis (namely~\cite{KKL88}, following~\cite{KN06}) for the same purpose, as well as~\cite{CKKRS06,CK07-hardness} for computational hardness. The best impossibility result currently known~\cite{KM13} for Sparsest Cut with Uniform Demands relies on the development of new  pseudorandom generators.

The idea of using the  geometry of the Heisenberg group to bound $\rho_{\mathsf{GL}}(n)$ from below originated in~\cite{LN06}, where the relevant metric of negative type was constructed through a complex-analytic argument, and initial (qualitative) impossibility results were presented through the use of Pansu's differentiation theorem~\cite{Pan89} and  the Radon--Nikod\'ym Property from functional analysis (see e.g.~\cite{BL00}). In~\cite{CK10}, it was shown that the Heisenberg group indeed provides a proof that $\lim_{n\to \infty} \rho_{\mathsf{GL}}(n)=\infty$.  This proof introduced a remarkable new notion of differentiation for $L_1$-valued mappings, which led to the use of tools from geometric measure theory~\cite{FSSCRectifiability,FSSC03} to study the problem. A different proof that $\H^3$ fails to admit a bi-Lipschitz embedding into $L_1$ was found in~\cite{CheegerKleinerMetricDiff}, where a classical notion of metric differentiation~\cite{Kir94} was used in conjunction with the novel idea to consider monotonicity of sets in this context, combined with a sub-Riemannian-geometric argument that yielded a classification of monotone subsets of $\H^3$.  The main result of~\cite{CKN} finds a quantitative lower estimate for the scale at which this differentiation argument can be applied, leading to a lower bound of $(\log n)^{\Omega(1)}$ on $\rho_{\mathsf{GL}}(n)$.  This result relies on a mixture of the methods of~\cite{CK10} and~\cite{CheegerKleinerMetricDiff} and requires overcoming obstacles that are not present in the original qualitative investigation. In particular, \cite{CKN} introduced the quantitative measures of non-monotonicity that we use in the present work to find crucial bounds in the construction of an intrinsic corona decomposition. The quantitative differentiation bound of~\cite{CKN} remains the best bound currently known, and it would be very interesting to discover the sharp behavior in this more subtle question.

The desire to avoid the (often difficult) need to obtain sharp bounds for quantitative differentiation motivated the investigations in~\cite{AusNaoTes,LafforgueNaor}. In particular, \cite{AusNaoTes} devised a method to prove sharp (up to lower order factors) nonembeddability statements for the Heisenberg group based on a cohomological argument and a quantitative ergodic theorem. For Hilbert-space valued mappings, \cite{AusNaoTes} used a cohomological argument in combination with representation theory to prove the following quadratic inequality for every finitely supported function $\upphi:\H_Z^5\to L_2$.
\begin{multline}\label{eq:discrete global quadratic}
\bigg(\sum_{t=1}^\infty \frac{1}{t^2}\sum_{h\in \Z^5} \big\|\upphi\big(hZ^t\big)-\upphi(h)\big\|_2^2\bigg)^{\frac12}\\\lesssim \bigg(\sum_{h\in \Z^5}\sum_{\sigma\in S}\big\|\upphi(h\sigma)-\upphi(h)\big\|_2^2\bigg)^{\frac12}.
\end{multline}
In~\cite{LafforgueNaor} a different approach based on Littlewood--Paley theory was devised, leading to the following generalization of~\eqref{eq:discrete global quadratic} that holds true for every $p\in (1,2]$ and every finitely supported $\upphi:\H^5\to L_p$.
\begin{multline}\label{eq:discrete global p}
\left(\sum_{t=1}^\infty \frac{1}{t^2}\bigg(\sum_{h\in \Z^5} \big\|\upphi\big(hZ^t\big)-\upphi(h)\big\|_p^p\bigg)^\frac{2}{p}\right)^{\frac12}\\\le  C(p)\bigg(\sum_{h\in \Z^5}\sum_{\sigma\in S}\big\|\upphi(h\sigma)-\upphi(h)\big\|_p^p\bigg)^{\frac{1}{p}},
\end{multline}
for some $C(p)\in (0,\infty)$. See~\cite{LafforgueNaor} for a strengthening of~\eqref{eq:discrete global p} that holds for general uniformly convex targets (using the recently established~\cite{MTX06} vector-valued Littlewood--Paley--Stein theory for the Poisson semigroup). These functional inequalities yield sharp non-embeddability estimates for balls in $\H^5_\Z$, but the method of~\cite{LafforgueNaor} inherently yields a constant $C(p)$ in~\eqref{eq:discrete global p} that satisfies $\lim_{p\to 1} C(p)=\infty$.  The estimate~\eqref{eq:discrete local intro} that we prove here for $L_1$-valued mappings is an endpoint estimate corresponding to~\eqref{eq:discrete global p}, showing that the best possible $C(p)$ actually remains bounded as $p\to 1$. This confirms a conjecture of~\cite{LafforgueNaor} and is crucial for the results that we obtain here.

As explained in Section~\ref{sec:into lip graph}, our proof of~\eqref{eq:discrete global quadratic} uses the $\H^3$-analogue of~\eqref{eq:discrete global quadratic}. It should be mentioned at this juncture that the proofs of~\eqref{eq:discrete global quadratic}  and~\eqref{eq:discrete global p} in~\cite{AusNaoTes,LafforgueNaor} were oblivious to the dimension of the underlying Heisenberg group.\footnote{Thus far in this extended abstract we recalled the definitions of $\H^5$ and $\H^3$ but not of higher-dimensional Heisenberg groups (since they are not needed for any of the applications that are  obtain here). Nevertheless, it is obvious how to generalize either the matrix group or the group modelled on $\R^5$ that we considered above to obtain the Heisenberg group $\H^{2k+1}$ for any $k\in \N$.} An unexpected aspect of the present work is that the underlying dimension does play a role at the endpoint $p=1$, with the analogue of~\eqref{eq:discrete local intro}  (or Theorem~\ref{thm:isoperimetric discrete}) for $\H^3$ being in fact {\em incorrect}; see Section~\ref{sec:3d} below. In the full version \cite{NY-full} of this paper we shall establish the $\H^{2k+1}$-analogue of Theorem~\ref{thm:isoperimetric discrete} for every $k\in \{2,3,\ldots\}$, in which case the implicit constant depends on $k$, and we shall also obtain the sharp asymptotic behavior as $k\to \infty$.

As we recalled above, past progress on the Sparsest Cut Problem   came hand-in-hand with meaningful mathematical developments. The present work is a culmination of a long-term project that is rooted in mathematical phenomena that are interesting not just for their relevance to approximation algorithms but also for their connections to the broader mathematical world.  In the ensuing subsections we shall describe some further results and questions related to this general direction.

\subsubsection{The $3$-dimensional Heisenberg group}\label{sec:3d} The investigation of the possible validity of an appropriate analogue of Theorem~\ref{thm:isoperimetric discrete} with $\H_\Z^5$ replaced by $\H_\Z^3$ remains an intriguing mystery and a subject of ongoing research that will be published elsewhere. This ongoing work shows that  Theorem~\ref{thm:isoperimetric discrete} fails for $\H_\Z^3$, but that there exists $p\in (2,\infty)$ such that for every $\Omega\subset \H_\Z^3$ we have
\begin{equation}\label{eq:p version}
\bigg(\sum_{t=1}^\infty \frac{|\partial^t_{\vv}\Omega|^p}{t^{1+\frac{p}{2}}}\bigg)^{\frac{1}{p}}\lesssim |\partial_{\hh}\Omega|.
\end{equation}
A simple argument shows that $\sup_{s\in \N} |\partial^s_{\vv}\Omega|/\sqrt{s}\le  \gamma |\partial_{\hh}\Omega|$ for some universal constant $\gamma>0$. Hence, for every $t\in \N$ we have \begin{align*}
\frac{|\partial^t_{\vv}\Omega|^p}{t^{1+\frac{p}{2}}}\le\frac{|\partial^t_{\vv}\Omega|^2}{t^2}\sup_{s\in \N} \left(\frac{|\partial^s_{\vv}\Omega|}{\sqrt{s}}\right)^{p-2} \le \frac{|\partial^t_{\vv}\Omega|^2}{t^2}(\gamma|\partial_{\hh}\Omega|)^{p-2}.
 \end{align*}This implies that the left hand side of~\eqref{eq:p version} is bounded from above by a universal  constant multiple of  $|\partial_{\vv}\Omega|^{2/p}|\partial_{\hh}\Omega|^{1-2/p}$. Therefore~\eqref{eq:p version} is  weaker than the estimate  $|\partial_{\vv}\Omega|\lesssim |\partial_{\hh}\Omega|$ of Theorem~\ref{thm:isoperimetric discrete}. It would be interesting to determine the infimum over those $p$ for which~\eqref{eq:p version} holds true for every $\Omega\subset \H_\Z^3$, with our ongoing work showing that it is at least $4$. In fact, this work shows that for every $R\ge 2$ the $L_1$ distortion of the ball of radius $R$ in $\H_\Z^3$ is at most a constant multiple of $\sqrt[4]{\log R}$ --- asymptotically \emph{less} than the distortion of the ball of the same radius in $\H_\Z^5$.  It would be interesting to determine the correct asymptotics of this distortion, with the best-known lower bound remaining that of~\cite{CKN}, i.e., a constant multiple of $(\log R)^\delta$  for some universal constant $\delta>0$. It should be stressed, however, that the algorithmic application of Theorem~\ref{thm:isoperimetric discrete} that is obtained here    uses Theorem~\eqref{thm:isoperimetric discrete} as stated for $\H_\Z^5$, and understanding the case of $\H_\Z^3$ would not yield any further improvement. So, while the above questions are  geometrically and analytically interesting in their own right,  they are not needed for applications that we currently have in mind.

\subsubsection{Metric embeddings}\label{sec:embed intro} Theorem~\ref{thm:distortion R} also yields a sharp result for the general problem of finding the asymptotically largest-possible $L_1$ distortion of a finite doubling metric space with $n$ points. A metric space $(X,d_X)$ is said to be $K$-doubling for some $K\in \N$ if every ball in $X$ (centered anywhere and of any radius) can be covered by $K$ balls of half its radius. By~\cite{KLMN05},
\begin{equation}\label{eq:descent}
c_1(X,d_X)\lesssim \sqrt{(\log K)\log |X|}.
\end{equation}
As noted in~\cite{GKL03}, the dependence on $|X|$ in~\eqref{eq:descent}, but with a worse dependence on $K$, follows by combining results of~\cite{Ass83} and~\cite{Rao99} (the dependence on $K$ that follows from~\cite{Ass83,Rao99}  was improved significantly in~\cite{GKL03}).  The metric space $(\Z^5,d_W)$ is $O(1)$-doubling because $|\mathcal{B}_{R}|\asymp R^6$ for every $R\ge 1$. Theorem~\ref{thm:distortion R} shows that~\eqref{eq:descent} is sharp  when $K=O(1)$, thus improving over the previously best-known construction~\cite{LS11} of arbitrarily large $O(1)$-doubling finite metric spaces $\{(X_i,d_i)\}_{i=1}^\infty$ for which $c_1(X_i,d_i)\gtrsim \sqrt{(\log |X_i|)/\log\log |X_i|}$. Probably~\eqref{eq:descent} is sharp for every $K\le |X|$; conceivably this could be proven by incorporating Theorem~\ref{thm:distortion R}  into the argument of~\cite{JLM11}, but we shall not pursue this here. Theorem~\ref{thm:distortion R} establishes for the first time the existence of a metric space that simultaneously has several useful geometric properties and poor (indeed, worst possible) embeddability into $L_1$. By virtue of being $O(1)$-doubling, the metric space $(\Z^5,d_W)$ also has Markov type $2$ due to~\cite{DLP13} (which improves over~\cite{NPSS06}, where the conclusion that it has Markov type $p$ for every $p<2$ was obtained). For more on the bi-Lipschitz invariant Markov type and its applications, see~\cite{Bal92,Nao12}. The property of having Markov type $2$ is  shared by the construction of~\cite{LS11}, which is also $O(1)$-doubling, but $(\Z^5,d_W)$ has additional features that  the example of~\cite{LS11} fails to have. For one, it is a group; for another, by~\cite{Li14,Li16} we know that $(\Z^5,d_W)$ has Markov convexity $4$ (and no less). (See~\cite{LNP09,MN13} for background on the bi-Lipschitz invariant Markov convexity and its consequences.) By~\cite[Section~3]{MN13} the example of~\cite{LS11} does not have Markov convexity $p$ for any finite $p$. No examples of arbitrarily large finite metric spaces $\{(X_i,d_i)\}_{i=1}^\infty$ with bounded Markov convexity (and Markov convexity constants uniformly bounded) such that $c_1(X_i,d_i)\gtrsim\sqrt{\log |X_i|}$ were previously known to exist. Analogous statements are known to be impossible for Banach spaces~\cite{MW78}, so it is natural in the context of the Ribe program (see the surveys~\cite{Nao12,Bal13} for more on this research program) to ask whether there is a potential metric version of~\cite{MW78}; the above discussion shows that there is not.

\subsubsection{The Sparsest Cut Problem with  Uniform Demands}\label{sec:uniform} An important special case of the Sparsest Cut Problem is when the demand matrix $D$ is the matrix $\1_{\n\times\n}\in M_n(\R)$ all of whose entries equal $1$ and the capacity matrix $C$ lies in $M_n(\{0,1\})$, i.e., all its entries are either $0$ or $1$. This is known as the {\em Sparsest Cut Problem with  Uniform Demands}. In this case $C$ can also be described as the adjacency matrix of a graph $G$ whose vertex set is $\n$ and whose edge set consists of those unordered pairs $\{i,j\}\subset \n$ for which $C_{ij}=1$. With this interpretation, given $A\subset \n$ the numerator in~\eqref{eq:def opt} equals twice the number of edges that are incident to $A$ in $G$. And, since  $D=\1_{\n\times\n}$, the  denominator in~\eqref{eq:def opt} is equal to $2|A|(n-|A|)\asymp n\min\{|A|,|\n\setminus A|\}$. So, the Sparsest Cut Problem with Uniform Demands asks for an algorithm that takes as input a finite graph and outputs a quantity which is bounded above and below by universal constant multiples of its {\em conductance}~\cite{JS89} divided by $n$. The Goemans--Linial integrality gap corresponding to this special case is
$$
\uprho_{\mathsf{GL}}^{\mathrm{unif}}(n)\eqdef \sup_{\substack{C\in M_n(\{0,1\})\\ C\ \mathrm{symmetric}}} \frac{\mathsf{OPT}(C,\1_{\n\times\n})}{\mathsf{SDP}(C,\1_{\n\times \n})}.
$$
The Goemans--Linial algorithm furnishes the best-known approximation ratio also in the  case of uniform demands. By the important work~\cite{ARV09} we have $\uprho_{\mathsf{GL}}^{\mathrm{unif}}(n)\lesssim \sqrt{\log n}$, improving over the previous bound $\uprho_{\mathsf{GL}}^{\mathrm{unif}}(n)\lesssim \log n$ of~\cite{LR99}. As explained in~\cite{CKN09}, the present approach based on (fixed dimensional) Heisenberg groups cannot yield a lower bound on $\uprho_{\mathsf{GL}}^{\mathrm{unif}}(n)$ that tends to $\infty$ with $n$. The best-known lower bound~\cite{KM13} is $\uprho_{\mathsf{GL}}^{\mathrm{unif}}(n)\ge \exp(c\sqrt{\log\log n})$ for some universal constant $c>0$, improving over the previous bound $\uprho_{\mathsf{GL}}^{\mathrm{unif}}(n)\gtrsim \log\log n$ of~\cite{DKSV06}.   Determining the asymptotic behavior of $\uprho_{\mathsf{GL}}^{\mathrm{unif}}(n)$ remains an intriguing open problem.

\bibliographystyle{ACM-Reference-Format}
\bibliography{corona,appendix}


\def\cprime{$'$} \def\cprime{$'$} \def\cprime{$'$} \def\cprime{$'$}
\begin{thebibliography}{00}

%%% the Web address.

\ifx \showCODEN    \undefined \def \showCODEN     #1{\unskip}     \fi
\ifx \showDOI      \undefined \def \showDOI       #1{{\tt DOI:}\penalty0{#1}\ }
  \fi
\ifx \showISBNx    \undefined \def \showISBNx     #1{\unskip}     \fi
\ifx \showISBNxiii \undefined \def \showISBNxiii  #1{\unskip}     \fi
\ifx \showISSN     \undefined \def \showISSN      #1{\unskip}     \fi
\ifx \showLCCN     \undefined \def \showLCCN      #1{\unskip}     \fi
\ifx \shownote     \undefined \def \shownote      #1{#1}          \fi
\ifx \showarticletitle \undefined \def \showarticletitle #1{#1}   \fi
\ifx \showURL      \undefined \def \showURL       #1{#1}          \fi
% invisible to TeX
\providecommand\bibfield[2]{#2}
\providecommand\bibinfo[2]{#2}
\providecommand\natexlab[1]{#1}
\providecommand\showeprint[2][]{arXiv:#2}

\bibitem[\protect\citeauthoryear{Agrawal, Klein, Ravi, and Rao}{Agrawal
  et~al\mbox{.}}{1990}]%
        {AKRR90}
\bibfield{author}{\bibinfo{person}{A. Agrawal}, \bibinfo{person}{P. Klein},
  \bibinfo{person}{R. Ravi}, {and} \bibinfo{person}{S. Rao}.}
  \bibinfo{year}{1990}\natexlab{}.
\newblock \showarticletitle{Approximation through multicommodity flow}.
\newblock In \bibinfo{booktitle}{{\em 31st Annual Symposium on Foundations of
  Computer Science}}. \bibinfo{publisher}{IEEE Computer Soc., Los Alamitos,
  CA}, \bibinfo{pages}{726--737}.
\newblock


\bibitem[\protect\citeauthoryear{Ambrosio}{Ambrosio}{2001}]%
        {Amb01}
\bibfield{author}{\bibinfo{person}{Luigi Ambrosio}.}
  \bibinfo{year}{2001}\natexlab{}.
\newblock \showarticletitle{Some fine properties of sets of finite perimeter in
  {A}hlfors regular metric measure spaces}.
\newblock \bibinfo{journal}{{\em Adv. Math.\/}} \bibinfo{volume}{159},
  \bibinfo{number}{1} (\bibinfo{year}{2001}), \bibinfo{pages}{51--67}.
\newblock
\showCODEN{ADMTA4}
\showISSN{0001-8708}
\showDOI{%
\url{http://dx.doi.org/10.1006/aima.2000.1963}}


\bibitem[\protect\citeauthoryear{Arora, Lee, and Naor}{Arora
  et~al\mbox{.}}{2008}]%
        {ALN08}
\bibfield{author}{\bibinfo{person}{Sanjeev Arora}, \bibinfo{person}{James~R.
  Lee}, {and} \bibinfo{person}{Assaf Naor}.} \bibinfo{year}{2008}\natexlab{}.
\newblock \showarticletitle{Euclidean distortion and the sparsest cut}.
\newblock \bibinfo{journal}{{\em J. Amer. Math. Soc.\/}} \bibinfo{volume}{21},
  \bibinfo{number}{1} (\bibinfo{year}{2008}), \bibinfo{pages}{1--21
  (electronic)}.
\newblock
\showISSN{0894-0347}
\showDOI{%
\url{http://dx.doi.org/10.1090/S0894-0347-07-00573-5}}


\bibitem[\protect\citeauthoryear{Arora, Rao, and Vazirani}{Arora
  et~al\mbox{.}}{2009}]%
        {ARV09}
\bibfield{author}{\bibinfo{person}{Sanjeev Arora}, \bibinfo{person}{Satish
  Rao}, {and} \bibinfo{person}{Umesh Vazirani}.}
  \bibinfo{year}{2009}\natexlab{}.
\newblock \showarticletitle{Expander flows, geometric embeddings and graph
  partitioning}.
\newblock \bibinfo{journal}{{\em J. ACM\/}} \bibinfo{volume}{56},
  \bibinfo{number}{2} (\bibinfo{year}{2009}), \bibinfo{pages}{Art. 5, 37}.
\newblock
\showISSN{0004-5411}
\showDOI{%
\url{http://dx.doi.org/10.1145/1502793.1502794}}


\bibitem[\protect\citeauthoryear{Assouad}{Assouad}{1983}]%
        {Ass83}
\bibfield{author}{\bibinfo{person}{Patrice Assouad}.}
  \bibinfo{year}{1983}\natexlab{}.
\newblock \showarticletitle{Plongements lipschitziens dans {${\bf R}^{n}$}}.
\newblock \bibinfo{journal}{{\em Bull. Soc. Math. France\/}}
  \bibinfo{volume}{111}, \bibinfo{number}{4} (\bibinfo{year}{1983}),
  \bibinfo{pages}{429--448}.
\newblock
\showCODEN{BSMFAA}
\showISSN{0037-9484}
\showURL{%
\url{http://www.numdam.org/item?id=BSMF_1983__111__429_0}}


\bibitem[\protect\citeauthoryear{Aumann and Rabani}{Aumann and Rabani}{1998}]%
        {AR98}
\bibfield{author}{\bibinfo{person}{Yonatan Aumann} {and} \bibinfo{person}{Yuval
  Rabani}.} \bibinfo{year}{1998}\natexlab{}.
\newblock \showarticletitle{An {$O(\log k)$} approximate min-cut max-flow
  theorem and approximation algorithm}.
\newblock \bibinfo{journal}{{\em SIAM J. Comput.\/}} \bibinfo{volume}{27},
  \bibinfo{number}{1} (\bibinfo{year}{1998}), \bibinfo{pages}{291--301
  (electronic)}.
\newblock
\showISSN{0097-5397}
\showDOI{%
\url{http://dx.doi.org/10.1137/S0097539794285983}}


\bibitem[\protect\citeauthoryear{Austin, Naor, and Tessera}{Austin
  et~al\mbox{.}}{2013}]%
        {AusNaoTes}
\bibfield{author}{\bibinfo{person}{Tim Austin}, \bibinfo{person}{Assaf Naor},
  {and} \bibinfo{person}{Romain Tessera}.} \bibinfo{year}{2013}\natexlab{}.
\newblock \showarticletitle{Sharp quantitative nonembeddability of the
  {H}eisenberg group into superreflexive {B}anach spaces}.
\newblock \bibinfo{journal}{{\em Groups Geom. Dyn.\/}} \bibinfo{volume}{7},
  \bibinfo{number}{3} (\bibinfo{year}{2013}), \bibinfo{pages}{497--522}.
\newblock
\showISSN{1661-7207}
\showDOI{%
\url{http://dx.doi.org/10.4171/GGD/193}}


\bibitem[\protect\citeauthoryear{Ball}{Ball}{1992}]%
        {Bal92}
\bibfield{author}{\bibinfo{person}{K. Ball}.} \bibinfo{year}{1992}\natexlab{}.
\newblock \showarticletitle{Markov chains, {R}iesz transforms and {L}ipschitz
  maps}.
\newblock \bibinfo{journal}{{\em Geom. Funct. Anal.\/}} \bibinfo{volume}{2},
  \bibinfo{number}{2} (\bibinfo{year}{1992}), \bibinfo{pages}{137--172}.
\newblock
\showCODEN{GFANFB}
\showISSN{1016-443X}
\showDOI{%
\url{http://dx.doi.org/10.1007/BF01896971}}


\bibitem[\protect\citeauthoryear{Ball}{Ball}{2013}]%
        {Bal13}
\bibfield{author}{\bibinfo{person}{Keith Ball}.}
  \bibinfo{year}{2013}\natexlab{}.
\newblock \showarticletitle{The {R}ibe programme}.
\newblock \bibinfo{journal}{{\em Ast\'erisque\/}} \bibinfo{number}{352}
  (\bibinfo{year}{2013}), \bibinfo{pages}{Exp. No. 1047, viii, 147--159}.
\newblock
\showISBNx{978-2-85629-371-3}
\showISSN{0303-1179}
\newblock
\shownote{S{\'e}minaire Bourbaki. Vol. 2011/2012. Expos{\'e}s 1043--1058.}


\bibitem[\protect\citeauthoryear{Bass}{Bass}{1972}]%
        {Bas72}
\bibfield{author}{\bibinfo{person}{H. Bass}.} \bibinfo{year}{1972}\natexlab{}.
\newblock \showarticletitle{The degree of polynomial growth of finitely
  generated nilpotent groups}.
\newblock \bibinfo{journal}{{\em Proc. London Math. Soc. (3)\/}}
  \bibinfo{volume}{25} (\bibinfo{year}{1972}), \bibinfo{pages}{603--614}.
\newblock
\showISSN{0024-6115}


\bibitem[\protect\citeauthoryear{Benyamini and Lindenstrauss}{Benyamini and
  Lindenstrauss}{2000}]%
        {BL00}
\bibfield{author}{\bibinfo{person}{Yoav Benyamini} {and} \bibinfo{person}{Joram
  Lindenstrauss}.} \bibinfo{year}{2000}\natexlab{}.
\newblock \bibinfo{booktitle}{{\em Geometric nonlinear functional analysis.
  {V}ol. 1}}. \bibinfo{series}{American Mathematical Society Colloquium
  Publications}, Vol.~\bibinfo{volume}{48}.
\newblock \bibinfo{publisher}{American Mathematical Society, Providence, RI}.
  xii+488 pages.
\newblock
\showISBNx{0-8218-0835-4}


\bibitem[\protect\citeauthoryear{Bourgain}{Bourgain}{1985}]%
        {Bou85}
\bibfield{author}{\bibinfo{person}{J. Bourgain}.}
  \bibinfo{year}{1985}\natexlab{}.
\newblock \showarticletitle{On {L}ipschitz embedding of finite metric spaces in
  {H}ilbert space}.
\newblock \bibinfo{journal}{{\em Israel J. Math.\/}} \bibinfo{volume}{52},
  \bibinfo{number}{1-2} (\bibinfo{year}{1985}), \bibinfo{pages}{46--52}.
\newblock
\showCODEN{ISJMAP}
\showISSN{0021-2172}
\showDOI{%
\url{http://dx.doi.org/10.1007/BF02776078}}


\bibitem[\protect\citeauthoryear{Bourgain}{Bourgain}{2002}]%
        {Bou02}
\bibfield{author}{\bibinfo{person}{J. Bourgain}.}
  \bibinfo{year}{2002}\natexlab{}.
\newblock \showarticletitle{On the distributions of the {F}ourier spectrum of
  {B}oolean functions}.
\newblock \bibinfo{journal}{{\em Israel J. Math.\/}}  \bibinfo{volume}{131}
  (\bibinfo{year}{2002}), \bibinfo{pages}{269--276}.
\newblock
\showCODEN{ISJMAP}
\showISSN{0021-2172}
\showDOI{%
\url{http://dx.doi.org/10.1007/BF02785861}}


\bibitem[\protect\citeauthoryear{Burago, Burago, and Ivanov}{Burago
  et~al\mbox{.}}{2001}]%
        {BBI01}
\bibfield{author}{\bibinfo{person}{Dmitri Burago}, \bibinfo{person}{Yuri
  Burago}, {and} \bibinfo{person}{Sergei Ivanov}.}
  \bibinfo{year}{2001}\natexlab{}.
\newblock \bibinfo{booktitle}{{\em A course in metric geometry}}.
  \bibinfo{series}{Graduate Studies in Mathematics}, Vol.~\bibinfo{volume}{33}.
\newblock \bibinfo{publisher}{American Mathematical Society, Providence, RI}.
  xiv+415 pages.
\newblock
\showISBNx{0-8218-2129-6}
\showDOI{%
\url{http://dx.doi.org/10.1090/gsm/033}}


\bibitem[\protect\citeauthoryear{Capogna, Danielli, Pauls, and Tyson}{Capogna
  et~al\mbox{.}}{2007}]%
        {CDPT07}
\bibfield{author}{\bibinfo{person}{Luca Capogna}, \bibinfo{person}{Donatella
  Danielli}, \bibinfo{person}{Scott~D. Pauls}, {and} \bibinfo{person}{Jeremy~T.
  Tyson}.} \bibinfo{year}{2007}\natexlab{}.
\newblock \bibinfo{booktitle}{{\em An introduction to the {H}eisenberg group
  and the sub-{R}iemannian isoperimetric problem}}. \bibinfo{series}{Progress
  in Mathematics}, Vol.~\bibinfo{volume}{259}.
\newblock \bibinfo{publisher}{Birkh\"auser Verlag, Basel}. xvi+223 pages.
\newblock
\showISBNx{978-3-7643-8132-5; 3-7643-8132-9}


\bibitem[\protect\citeauthoryear{Chawla}{Chawla}{2008}]%
        {Chawla08}
\bibfield{author}{\bibinfo{person}{Shuchi Chawla}.}
  \bibinfo{year}{2008}\natexlab{}.
\newblock \showarticletitle{Sparsest Cut}.
\newblock In \bibinfo{booktitle}{{\em Encyclopedia of Algorithms}}.
  \bibinfo{publisher}{Springer-Verlag US}, \bibinfo{pages}{868--870}.
\newblock


\bibitem[\protect\citeauthoryear{Chawla, Gupta, and R{\"a}cke}{Chawla
  et~al\mbox{.}}{2008}]%
        {CGR08}
\bibfield{author}{\bibinfo{person}{Shuchi Chawla}, \bibinfo{person}{Anupam
  Gupta}, {and} \bibinfo{person}{Harald R{\"a}cke}.}
  \bibinfo{year}{2008}\natexlab{}.
\newblock \showarticletitle{Embeddings of negative-type metrics and an improved
  approximation to generalized sparsest cut}.
\newblock \bibinfo{journal}{{\em ACM Trans. Algorithms\/}} \bibinfo{volume}{4},
  \bibinfo{number}{2} (\bibinfo{year}{2008}), \bibinfo{pages}{Art. 22, 18}.
\newblock
\showISSN{1549-6325}
\showDOI{%
\url{http://dx.doi.org/10.1145/1361192.1361199}}


\bibitem[\protect\citeauthoryear{Chawla, Krauthgamer, Kumar, Rabani, and
  Sivakumar}{Chawla et~al\mbox{.}}{2006}]%
        {CKKRS06}
\bibfield{author}{\bibinfo{person}{Shuchi Chawla}, \bibinfo{person}{Robert
  Krauthgamer}, \bibinfo{person}{Ravi Kumar}, \bibinfo{person}{Yuval Rabani},
  {and} \bibinfo{person}{D. Sivakumar}.} \bibinfo{year}{2006}\natexlab{}.
\newblock \showarticletitle{On the hardness of approximating multicut and
  sparsest-cut}.
\newblock \bibinfo{journal}{{\em Comput. Complexity\/}} \bibinfo{volume}{15},
  \bibinfo{number}{2} (\bibinfo{year}{2006}), \bibinfo{pages}{94--114}.
\newblock
\showCODEN{CPTCEU}
\showISSN{1016-3328}
\showDOI{%
\url{http://dx.doi.org/10.1007/s00037-006-0210-9}}


\bibitem[\protect\citeauthoryear{Cheeger and Kleiner}{Cheeger and
  Kleiner}{2010a}]%
        {CK10}
\bibfield{author}{\bibinfo{person}{Jeff Cheeger} {and} \bibinfo{person}{Bruce
  Kleiner}.} \bibinfo{year}{2010}\natexlab{a}.
\newblock \showarticletitle{Differentiating maps into {$L^1$}, and the geometry
  of {BV} functions}.
\newblock \bibinfo{journal}{{\em Ann. of Math. (2)\/}} \bibinfo{volume}{171},
  \bibinfo{number}{2} (\bibinfo{year}{2010}), \bibinfo{pages}{1347--1385}.
\newblock
\showCODEN{ANMAAH}
\showISSN{0003-486X}
\showDOI{%
\url{http://dx.doi.org/10.4007/annals.2010.171.1347}}


\bibitem[\protect\citeauthoryear{Cheeger and Kleiner}{Cheeger and
  Kleiner}{2010b}]%
        {CheegerKleinerMetricDiff}
\bibfield{author}{\bibinfo{person}{Jeff Cheeger} {and} \bibinfo{person}{Bruce
  Kleiner}.} \bibinfo{year}{2010}\natexlab{b}.
\newblock \showarticletitle{Metric differentiation, monotonicity and maps to
  {$L\sp 1$}}.
\newblock \bibinfo{journal}{{\em Invent. Math.\/}} \bibinfo{volume}{182},
  \bibinfo{number}{2} (\bibinfo{year}{2010}), \bibinfo{pages}{335--370}.
\newblock
\showCODEN{INVMBH}
\showISSN{0020-9910}
\showDOI{%
\url{http://dx.doi.org/10.1007/s00222-010-0264-9}}


\bibitem[\protect\citeauthoryear{Cheeger, Kleiner, and Naor}{Cheeger
  et~al\mbox{.}}{2009}]%
        {CKN09}
\bibfield{author}{\bibinfo{person}{Jeff Cheeger}, \bibinfo{person}{Bruce
  Kleiner}, {and} \bibinfo{person}{Assaf Naor}.}
  \bibinfo{year}{2009}\natexlab{}.
\newblock \showarticletitle{A {$(\log n)^{\Omega(1)}$} integrality gap for the
  sparsest cut {SDP}}.
\newblock In \bibinfo{booktitle}{{\em 2009 50th {A}nnual {IEEE} {S}ymposium on
  {F}oundations of {C}omputer {S}cience ({FOCS} 2009)}}.
  \bibinfo{publisher}{IEEE Computer Soc., Los Alamitos, CA},
  \bibinfo{pages}{555--564}.
\newblock
\showDOI{%
\url{http://dx.doi.org/10.1109/FOCS.2009.47}}


\bibitem[\protect\citeauthoryear{Cheeger, Kleiner, and Naor}{Cheeger
  et~al\mbox{.}}{2011}]%
        {CKN}
\bibfield{author}{\bibinfo{person}{Jeff Cheeger}, \bibinfo{person}{Bruce
  Kleiner}, {and} \bibinfo{person}{Assaf Naor}.}
  \bibinfo{year}{2011}\natexlab{}.
\newblock \showarticletitle{Compression bounds for {L}ipschitz maps from the
  {H}eisenberg group to {$L\sb 1$}}.
\newblock \bibinfo{journal}{{\em Acta Math.\/}} \bibinfo{volume}{207},
  \bibinfo{number}{2} (\bibinfo{year}{2011}), \bibinfo{pages}{291--373}.
\newblock
\showCODEN{ACMAA8}
\showISSN{0001-5962}
\showDOI{%
\url{http://dx.doi.org/10.1007/s11511-012-0071-9}}


\bibitem[\protect\citeauthoryear{Christ}{Christ}{1990}]%
        {ChristTb}
\bibfield{author}{\bibinfo{person}{Michael Christ}.}
  \bibinfo{year}{1990}\natexlab{}.
\newblock \showarticletitle{A {$T(b)$} theorem with remarks on analytic
  capacity and the {C}auchy integral}.
\newblock \bibinfo{journal}{{\em Colloq. Math.\/}} \bibinfo{volume}{60/61},
  \bibinfo{number}{2} (\bibinfo{year}{1990}), \bibinfo{pages}{601--628}.
\newblock
\showCODEN{CQMAAQ}
\showISSN{0010-1354}


\bibitem[\protect\citeauthoryear{Chuzhoy and Khanna}{Chuzhoy and
  Khanna}{2007}]%
        {CK07-hardness}
\bibfield{author}{\bibinfo{person}{Julia Chuzhoy} {and}
  \bibinfo{person}{Sanjeev Khanna}.} \bibinfo{year}{2007}\natexlab{}.
\newblock \showarticletitle{Polynomial flow-cut gaps and hardness of directed
  cut problems [extended abstract]}.
\newblock In \bibinfo{booktitle}{{\em S{TOC}'07---{P}roceedings of the 39th
  {A}nnual {ACM} {S}ymposium on {T}heory of {C}omputing}}.
  \bibinfo{publisher}{ACM}, \bibinfo{address}{New York},
  \bibinfo{pages}{179--188}.
\newblock


\bibitem[\protect\citeauthoryear{Chuzhoy and Khanna}{Chuzhoy and
  Khanna}{2009}]%
        {CK09-hardness}
\bibfield{author}{\bibinfo{person}{Julia Chuzhoy} {and}
  \bibinfo{person}{Sanjeev Khanna}.} \bibinfo{year}{2009}\natexlab{}.
\newblock \showarticletitle{Polynomial flow-cut gaps and hardness of directed
  cut problems}.
\newblock \bibinfo{journal}{{\em J. ACM\/}} \bibinfo{volume}{56},
  \bibinfo{number}{2} (\bibinfo{year}{2009}), \bibinfo{pages}{Art. 6, 28}.
\newblock
\showISSN{0004-5411}
\showDOI{%
\url{http://dx.doi.org/10.1145/1502793.1502795}}


\bibitem[\protect\citeauthoryear{David}{David}{1984}]%
        {Dav84}
\bibfield{author}{\bibinfo{person}{Guy David}.}
  \bibinfo{year}{1984}\natexlab{}.
\newblock \showarticletitle{Op\'erateurs int\'egraux singuliers sur certaines
  courbes du plan complexe}.
\newblock \bibinfo{journal}{{\em Ann. Sci. \'Ecole Norm. Sup. (4)\/}}
  \bibinfo{volume}{17}, \bibinfo{number}{1} (\bibinfo{year}{1984}),
  \bibinfo{pages}{157--189}.
\newblock
\showCODEN{ASENAH}
\showISSN{0012-9593}
\showURL{%
\url{http://www.numdam.org/item?id=ASENS_1984_4_17_1_157_0}}


\bibitem[\protect\citeauthoryear{David}{David}{1991a}]%
        {Dav91}
\bibfield{author}{\bibinfo{person}{Guy David}.}
  \bibinfo{year}{1991}\natexlab{a}.
\newblock \bibinfo{booktitle}{{\em Wavelets and singular integrals on curves
  and surfaces}}. \bibinfo{series}{Lecture Notes in Mathematics},
  Vol.~\bibinfo{volume}{1465}.
\newblock \bibinfo{publisher}{Springer-Verlag, Berlin}. x+107 pages.
\newblock
\showISBNx{3-540-53902-6}
\showDOI{%
\url{http://dx.doi.org/10.1007/BFb0091544}}


\bibitem[\protect\citeauthoryear{David}{David}{1991b}]%
        {DavidWavelets}
\bibfield{author}{\bibinfo{person}{Guy David}.}
  \bibinfo{year}{1991}\natexlab{b}.
\newblock \bibinfo{booktitle}{{\em Wavelets and singular integrals on curves
  and surfaces}}. \bibinfo{series}{Lecture Notes in Mathematics},
  Vol.~\bibinfo{volume}{1465}.
\newblock \bibinfo{publisher}{Springer-Verlag, Berlin}. x+107 pages.
\newblock
\showISBNx{3-540-53902-6}
\showDOI{%
\url{http://dx.doi.org/10.1007/BFb0091544}}


\bibitem[\protect\citeauthoryear{David and Semmes}{David and Semmes}{1991}]%
        {DavidSemmesSingular}
\bibfield{author}{\bibinfo{person}{G. David} {and} \bibinfo{person}{S.
  Semmes}.} \bibinfo{year}{1991}\natexlab{}.
\newblock \showarticletitle{Singular integrals and rectifiable sets in {${\bf
  R}\sp n$}: {B}eyond {L}ipschitz graphs}.
\newblock \bibinfo{journal}{{\em Ast\'erisque\/}} \bibinfo{number}{193}
  (\bibinfo{year}{1991}), \bibinfo{pages}{152}.
\newblock
\showISSN{0303-1179}


\bibitem[\protect\citeauthoryear{David and Semmes}{David and Semmes}{1993}]%
        {DSAnalysis}
\bibfield{author}{\bibinfo{person}{Guy David} {and} \bibinfo{person}{Stephen
  Semmes}.} \bibinfo{year}{1993}\natexlab{}.
\newblock \bibinfo{booktitle}{{\em Analysis of and on uniformly rectifiable
  sets}}. \bibinfo{series}{Mathematical Surveys and Monographs},
  Vol.~\bibinfo{volume}{38}.
\newblock \bibinfo{publisher}{American Mathematical Society, Providence, RI}.
  xii+356 pages.
\newblock
\showISBNx{0-8218-1537-7}
\showDOI{%
\url{http://dx.doi.org/10.1090/surv/038}}


\bibitem[\protect\citeauthoryear{Devanur, Khot, Saket, and Vishnoi}{Devanur
  et~al\mbox{.}}{2006}]%
        {DKSV06}
\bibfield{author}{\bibinfo{person}{Nikhil~R. Devanur},
  \bibinfo{person}{Subhash~A. Khot}, \bibinfo{person}{Rishi Saket}, {and}
  \bibinfo{person}{Nisheeth~K. Vishnoi}.} \bibinfo{year}{2006}\natexlab{}.
\newblock \showarticletitle{Integrality gaps for sparsest cut and minimum
  linear arrangement problems}.
\newblock In \bibinfo{booktitle}{{\em S{TOC}'06: {P}roceedings of the 38th
  {A}nnual {ACM} {S}ymposium on {T}heory of {C}omputing}}.
  \bibinfo{publisher}{ACM, New York}, \bibinfo{pages}{537--546}.
\newblock
\showDOI{%
\url{http://dx.doi.org/10.1145/1132516.1132594}}


\bibitem[\protect\citeauthoryear{Deza and Laurent}{Deza and Laurent}{1997}]%
        {DL97}
\bibfield{author}{\bibinfo{person}{Michel~Marie Deza} {and}
  \bibinfo{person}{Monique Laurent}.} \bibinfo{year}{1997}\natexlab{}.
\newblock \bibinfo{booktitle}{{\em Geometry of cuts and metrics}}.
  \bibinfo{series}{Algorithms and Combinatorics}, Vol.~\bibinfo{volume}{15}.
\newblock \bibinfo{publisher}{Springer-Verlag, Berlin}. xii+587 pages.
\newblock
\showISBNx{3-540-61611-X}
\showDOI{%
\url{http://dx.doi.org/10.1007/978-3-642-04295-9}}


\bibitem[\protect\citeauthoryear{Ding, Lee, and Peres}{Ding
  et~al\mbox{.}}{2013}]%
        {DLP13}
\bibfield{author}{\bibinfo{person}{Jian Ding}, \bibinfo{person}{James~R. Lee},
  {and} \bibinfo{person}{Yuval Peres}.} \bibinfo{year}{2013}\natexlab{}.
\newblock \showarticletitle{Markov type and threshold embeddings}.
\newblock \bibinfo{journal}{{\em Geom. Funct. Anal.\/}} \bibinfo{volume}{23},
  \bibinfo{number}{4} (\bibinfo{year}{2013}), \bibinfo{pages}{1207--1229}.
\newblock
\showISSN{1016-443X}
\showDOI{%
\url{http://dx.doi.org/10.1007/s00039-013-0234-7}}


\bibitem[\protect\citeauthoryear{Franchi, Serapioni, and Serra~Cassano}{Franchi
  et~al\mbox{.}}{2001}]%
        {FSSCRectifiability}
\bibfield{author}{\bibinfo{person}{Bruno Franchi}, \bibinfo{person}{Raul
  Serapioni}, {and} \bibinfo{person}{Francesco Serra~Cassano}.}
  \bibinfo{year}{2001}\natexlab{}.
\newblock \showarticletitle{Rectifiability and perimeter in the {H}eisenberg
  group}.
\newblock \bibinfo{journal}{{\em Math. Ann.\/}} \bibinfo{volume}{321},
  \bibinfo{number}{3} (\bibinfo{year}{2001}), \bibinfo{pages}{479--531}.
\newblock
\showCODEN{MAANA}
\showISSN{0025-5831}
\showDOI{%
\url{http://dx.doi.org/10.1007/s002080100228}}


\bibitem[\protect\citeauthoryear{Franchi, Serapioni, and Serra~Cassano}{Franchi
  et~al\mbox{.}}{2003}]%
        {FSSC03}
\bibfield{author}{\bibinfo{person}{Bruno Franchi}, \bibinfo{person}{Raul
  Serapioni}, {and} \bibinfo{person}{Francesco Serra~Cassano}.}
  \bibinfo{year}{2003}\natexlab{}.
\newblock \showarticletitle{On the structure of finite perimeter sets in step 2
  {C}arnot groups}.
\newblock \bibinfo{journal}{{\em J. Geom. Anal.\/}} \bibinfo{volume}{13},
  \bibinfo{number}{3} (\bibinfo{year}{2003}), \bibinfo{pages}{421--466}.
\newblock
\showISSN{1050-6926}
\showDOI{%
\url{http://dx.doi.org/10.1007/BF02922053}}


\bibitem[\protect\citeauthoryear{Franchi, Serapioni, and Serra~Cassano}{Franchi
  et~al\mbox{.}}{2006}]%
        {FSSC06}
\bibfield{author}{\bibinfo{person}{Bruno Franchi}, \bibinfo{person}{Raul
  Serapioni}, {and} \bibinfo{person}{Francesco Serra~Cassano}.}
  \bibinfo{year}{2006}\natexlab{}.
\newblock \showarticletitle{Intrinsic {L}ipschitz graphs in {H}eisenberg
  groups}.
\newblock \bibinfo{journal}{{\em J. Nonlinear Convex Anal.\/}}
  \bibinfo{volume}{7}, \bibinfo{number}{3} (\bibinfo{year}{2006}),
  \bibinfo{pages}{423--441}.
\newblock
\showISSN{1345-4773}


\bibitem[\protect\citeauthoryear{Franchi, Serapioni, and Serra~Cassano}{Franchi
  et~al\mbox{.}}{2011}]%
        {FSSCDifferentiability}
\bibfield{author}{\bibinfo{person}{Bruno Franchi}, \bibinfo{person}{Raul
  Serapioni}, {and} \bibinfo{person}{Francesco Serra~Cassano}.}
  \bibinfo{year}{2011}\natexlab{}.
\newblock \showarticletitle{Differentiability of intrinsic {L}ipschitz
  functions within {H}eisenberg groups}.
\newblock \bibinfo{journal}{{\em J. Geom. Anal.\/}} \bibinfo{volume}{21},
  \bibinfo{number}{4} (\bibinfo{year}{2011}), \bibinfo{pages}{1044--1084}.
\newblock
\showISSN{1050-6926}
\showDOI{%
\url{http://dx.doi.org/10.1007/s12220-010-9178-4}}


\bibitem[\protect\citeauthoryear{Goemans}{Goemans}{1997}]%
        {Goe97}
\bibfield{author}{\bibinfo{person}{Michel~X. Goemans}.}
  \bibinfo{year}{1997}\natexlab{}.
\newblock \showarticletitle{Semidefinite programming in combinatorial
  optimization}.
\newblock \bibinfo{journal}{{\em Math. Programming\/}} \bibinfo{volume}{79},
  \bibinfo{number}{1-3, Ser. B} (\bibinfo{year}{1997}),
  \bibinfo{pages}{143--161}.
\newblock
\showCODEN{MHPGA4}
\showISSN{0025-5610}
\newblock
\shownote{Lectures on mathematical programming (ismp97) (Lausanne, 1997).}


\bibitem[\protect\citeauthoryear{Gr{\"o}tschel, Lov{\'a}sz, and
  Schrijver}{Gr{\"o}tschel et~al\mbox{.}}{1993}]%
        {GLS93}
\bibfield{author}{\bibinfo{person}{Martin Gr{\"o}tschel},
  \bibinfo{person}{L{\'a}szl{\'o} Lov{\'a}sz}, {and} \bibinfo{person}{Alexander
  Schrijver}.} \bibinfo{year}{1993}\natexlab{}.
\newblock \bibinfo{booktitle}{{\em Geometric algorithms and combinatorial
  optimization\/} (\bibinfo{edition}{second} ed.)}. \bibinfo{series}{Algorithms
  and Combinatorics}, Vol.~\bibinfo{volume}{2}.
\newblock \bibinfo{publisher}{Springer-Verlag}, \bibinfo{address}{Berlin}.
  xii+362 pages.
\newblock
\showISBNx{3-540-56740-2}


\bibitem[\protect\citeauthoryear{Gupta, Krauthgamer, and Lee}{Gupta
  et~al\mbox{.}}{2003}]%
        {GKL03}
\bibfield{author}{\bibinfo{person}{Anupam Gupta}, \bibinfo{person}{Robert
  Krauthgamer}, {and} \bibinfo{person}{James~R. Lee}.}
  \bibinfo{year}{2003}\natexlab{}.
\newblock \showarticletitle{Bounded Geometries, Fractals, and Low-Distortion
  Embeddings}. In \bibinfo{booktitle}{{\em 44th Symposium on Foundations of
  Computer Science {(FOCS} 2003), 11-14 October 2003, Cambridge, MA, USA,
  Proceedings}}. \bibinfo{publisher}{{IEEE} Computer Society},
  \bibinfo{pages}{534--543}.
\newblock
\showISBNx{0-7695-2040-5}
\showDOI{%
\url{http://dx.doi.org/10.1109/SFCS.2003.1238226}}


\bibitem[\protect\citeauthoryear{Jaffe, Lee, and Moharrami}{Jaffe
  et~al\mbox{.}}{2011}]%
        {JLM11}
\bibfield{author}{\bibinfo{person}{Alexander Jaffe}, \bibinfo{person}{James~R.
  Lee}, {and} \bibinfo{person}{Mohammad Moharrami}.}
  \bibinfo{year}{2011}\natexlab{}.
\newblock \showarticletitle{On the optimality of gluing over scales}.
\newblock \bibinfo{journal}{{\em Discrete Comput. Geom.\/}}
  \bibinfo{volume}{46}, \bibinfo{number}{2} (\bibinfo{year}{2011}),
  \bibinfo{pages}{270--282}.
\newblock
\showCODEN{DCGEER}
\showISSN{0179-5376}
\showDOI{%
\url{http://dx.doi.org/10.1007/s00454-011-9359-3}}


\bibitem[\protect\citeauthoryear{Jones}{Jones}{1989}]%
        {Jon89}
\bibfield{author}{\bibinfo{person}{Peter~W. Jones}.}
  \bibinfo{year}{1989}\natexlab{}.
\newblock \showarticletitle{Square functions, {C}auchy integrals, analytic
  capacity, and harmonic measure}.
\newblock In \bibinfo{booktitle}{{\em Harmonic analysis and partial
  differential equations ({E}l {E}scorial, 1987)}}. \bibinfo{series}{Lecture
  Notes in Math.}, Vol.~\bibinfo{volume}{1384}. \bibinfo{publisher}{Springer,
  Berlin}, \bibinfo{pages}{24--68}.
\newblock
\showDOI{%
\url{http://dx.doi.org/10.1007/BFb0086793}}


\bibitem[\protect\citeauthoryear{Jones}{Jones}{1990}]%
        {Jon90}
\bibfield{author}{\bibinfo{person}{Peter~W. Jones}.}
  \bibinfo{year}{1990}\natexlab{}.
\newblock \showarticletitle{Rectifiable sets and the traveling salesman
  problem}.
\newblock \bibinfo{journal}{{\em Invent. Math.\/}} \bibinfo{volume}{102},
  \bibinfo{number}{1} (\bibinfo{year}{1990}), \bibinfo{pages}{1--15}.
\newblock
\showCODEN{INVMBH}
\showISSN{0020-9910}
\showDOI{%
\url{http://dx.doi.org/10.1007/BF01233418}}


\bibitem[\protect\citeauthoryear{Kahn, Kalai, and Linial}{Kahn
  et~al\mbox{.}}{1988}]%
        {KKL88}
\bibfield{author}{\bibinfo{person}{Jeff Kahn}, \bibinfo{person}{Gil Kalai},
  {and} \bibinfo{person}{Nathan Linial}.} \bibinfo{year}{1988}\natexlab{}.
\newblock \showarticletitle{The Influence of Variables on Boolean Functions
  (Extended Abstract)}. In \bibinfo{booktitle}{{\em 29th Annual Symposium on
  Foundations of Computer Science, White Plains, New York, USA, 24-26 October
  1988}}. \bibinfo{publisher}{{IEEE} Computer Society},
  \bibinfo{pages}{68--80}.
\newblock
\showISBNx{0-8186-0877-3}
\showDOI{%
\url{http://dx.doi.org/10.1109/SFCS.1988.21923}}


\bibitem[\protect\citeauthoryear{Kane and Meka}{Kane and Meka}{2013}]%
        {KM13}
\bibfield{author}{\bibinfo{person}{Daniel Kane} {and} \bibinfo{person}{Raghu
  Meka}.} \bibinfo{year}{2013}\natexlab{}.
\newblock \showarticletitle{A {PRG} for {L}ipschitz functions of polynomials
  with applications to sparsest cut}.
\newblock In \bibinfo{booktitle}{{\em S{TOC}'13---{P}roceedings of the 2013
  {ACM} {S}ymposium on {T}heory of {C}omputing}}. \bibinfo{publisher}{ACM, New
  York}, \bibinfo{pages}{1--10}.
\newblock
\showDOI{%
\url{http://dx.doi.org/10.1145/2488608.2488610}}


\bibitem[\protect\citeauthoryear{Khot}{Khot}{2002}]%
        {Kho02}
\bibfield{author}{\bibinfo{person}{Subhash Khot}.}
  \bibinfo{year}{2002}\natexlab{}.
\newblock \showarticletitle{On the power of unique 2-prover 1-round games}. In
  \bibinfo{booktitle}{{\em Proceedings of the {T}hirty-{F}ourth {A}nnual {ACM}
  {S}ymposium on {T}heory of {C}omputing}}. \bibinfo{publisher}{ACM, New York},
  \bibinfo{pages}{767--775}.
\newblock
\showDOI{%
\url{http://dx.doi.org/10.1145/509907.510017}}


\bibitem[\protect\citeauthoryear{Khot}{Khot}{2010}]%
        {Kho10}
\bibfield{author}{\bibinfo{person}{Subhash Khot}.}
  \bibinfo{year}{2010}\natexlab{}.
\newblock \showarticletitle{Inapproximability of {NP}-complete problems,
  discrete {F}ourier analysis, and geometry}. In \bibinfo{booktitle}{{\em
  Proceedings of the {I}nternational {C}ongress of {M}athematicians. {V}olume
  {IV}}}. \bibinfo{publisher}{Hindustan Book Agency, New Delhi},
  \bibinfo{pages}{2676--2697}.
\newblock


\bibitem[\protect\citeauthoryear{Khot and Naor}{Khot and Naor}{2006}]%
        {KN06}
\bibfield{author}{\bibinfo{person}{Subhash Khot} {and} \bibinfo{person}{Assaf
  Naor}.} \bibinfo{year}{2006}\natexlab{}.
\newblock \showarticletitle{Nonembeddability theorems via {F}ourier analysis}.
\newblock \bibinfo{journal}{{\em Math. Ann.\/}} \bibinfo{volume}{334},
  \bibinfo{number}{4} (\bibinfo{year}{2006}), \bibinfo{pages}{821--852}.
\newblock
\showCODEN{MAANA}
\showISSN{0025-5831}
\showDOI{%
\url{http://dx.doi.org/10.1007/s00208-005-0745-0}}


\bibitem[\protect\citeauthoryear{Khot and Vishnoi}{Khot and Vishnoi}{2015}]%
        {KV15}
\bibfield{author}{\bibinfo{person}{Subhash~A. Khot} {and}
  \bibinfo{person}{Nisheeth~K. Vishnoi}.} \bibinfo{year}{2015}\natexlab{}.
\newblock \showarticletitle{The unique games conjecture, integrability gap for
  cut problems and embeddability of negative-type metrics into {$\ell_1$}}.
\newblock \bibinfo{journal}{{\em J. ACM\/}} \bibinfo{volume}{62},
  \bibinfo{number}{1} (\bibinfo{year}{2015}), \bibinfo{pages}{Art. 8, 39}.
\newblock
\showISSN{0004-5411}
\showDOI{%
\url{http://dx.doi.org/10.1145/2629614}}


\bibitem[\protect\citeauthoryear{Kirchheim}{Kirchheim}{1994}]%
        {Kir94}
\bibfield{author}{\bibinfo{person}{Bernd Kirchheim}.}
  \bibinfo{year}{1994}\natexlab{}.
\newblock \showarticletitle{Rectifiable metric spaces: local structure and
  regularity of the {H}ausdorff measure}.
\newblock \bibinfo{journal}{{\em Proc. Amer. Math. Soc.\/}}
  \bibinfo{volume}{121}, \bibinfo{number}{1} (\bibinfo{year}{1994}),
  \bibinfo{pages}{113--123}.
\newblock
\showCODEN{PAMYAR}
\showISSN{0002-9939}
\showDOI{%
\url{http://dx.doi.org/10.2307/2160371}}


\bibitem[\protect\citeauthoryear{Krauthgamer, Lee, Mendel, and
  Naor}{Krauthgamer et~al\mbox{.}}{2005}]%
        {KLMN05}
\bibfield{author}{\bibinfo{person}{R. Krauthgamer}, \bibinfo{person}{J.~R.
  Lee}, \bibinfo{person}{M. Mendel}, {and} \bibinfo{person}{A. Naor}.}
  \bibinfo{year}{2005}\natexlab{}.
\newblock \showarticletitle{Measured descent: a new embedding method for finite
  metrics}.
\newblock \bibinfo{journal}{{\em Geom. Funct. Anal.\/}} \bibinfo{volume}{15},
  \bibinfo{number}{4} (\bibinfo{year}{2005}), \bibinfo{pages}{839--858}.
\newblock
\showCODEN{GFANFB}
\showISSN{1016-443X}
\showDOI{%
\url{http://dx.doi.org/10.1007/s00039-005-0527-6}}


\bibitem[\protect\citeauthoryear{Krauthgamer and Rabani}{Krauthgamer and
  Rabani}{2009}]%
        {KR09}
\bibfield{author}{\bibinfo{person}{Robert Krauthgamer} {and}
  \bibinfo{person}{Yuval Rabani}.} \bibinfo{year}{2009}\natexlab{}.
\newblock \showarticletitle{Improved lower bounds for embeddings into {$L_1$}}.
\newblock \bibinfo{journal}{{\em SIAM J. Comput.\/}} \bibinfo{volume}{38},
  \bibinfo{number}{6} (\bibinfo{year}{2009}), \bibinfo{pages}{2487--2498}.
\newblock
\showISSN{0097-5397}
\showDOI{%
\url{http://dx.doi.org/10.1137/060660126}}


\bibitem[\protect\citeauthoryear{Lafforgue and Naor}{Lafforgue and
  Naor}{2014}]%
        {LafforgueNaor}
\bibfield{author}{\bibinfo{person}{Vincent Lafforgue} {and}
  \bibinfo{person}{Assaf Naor}.} \bibinfo{year}{2014}\natexlab{}.
\newblock \showarticletitle{Vertical versus horizontal {P}oincar\'e
  inequalities on the {H}eisenberg group}.
\newblock \bibinfo{journal}{{\em Israel J. Math.\/}} \bibinfo{volume}{203},
  \bibinfo{number}{1} (\bibinfo{year}{2014}), \bibinfo{pages}{309--339}.
\newblock
\showISSN{0021-2172}
\showDOI{%
\url{http://dx.doi.org/10.1007/s11856-014-1088-x}}


\bibitem[\protect\citeauthoryear{Lee}{Lee}{2005}]%
        {Lee05}
\bibfield{author}{\bibinfo{person}{James~R. Lee}.}
  \bibinfo{year}{2005}\natexlab{}.
\newblock \showarticletitle{On distance scales, embeddings, and efficient
  relaxations of the cut cone}. In \bibinfo{booktitle}{{\em Proceedings of the
  {S}ixteenth {A}nnual {ACM}-{SIAM} {S}ymposium on {D}iscrete {A}lgorithms}}.
  \bibinfo{publisher}{ACM, New York}, \bibinfo{pages}{92--101 (electronic)}.
\newblock


\bibitem[\protect\citeauthoryear{Lee, Mendel, and Naor}{Lee
  et~al\mbox{.}}{2005}]%
        {LMN05}
\bibfield{author}{\bibinfo{person}{James~R. Lee}, \bibinfo{person}{Manor
  Mendel}, {and} \bibinfo{person}{Assaf Naor}.}
  \bibinfo{year}{2005}\natexlab{}.
\newblock \showarticletitle{Metric structures in {$L_1$}: dimension,
  snowflakes, and average distortion}.
\newblock \bibinfo{journal}{{\em European J. Combin.\/}} \bibinfo{volume}{26},
  \bibinfo{number}{8} (\bibinfo{year}{2005}), \bibinfo{pages}{1180--1190}.
\newblock
\showISSN{0195-6698}
\showDOI{%
\url{http://dx.doi.org/10.1016/j.ejc.2004.07.002}}


\bibitem[\protect\citeauthoryear{Lee and Naor}{Lee and Naor}{2006}]%
        {LN06}
\bibfield{author}{\bibinfo{person}{James~R. Lee} {and} \bibinfo{person}{Assaf
  Naor}.} \bibinfo{year}{2006}\natexlab{}.
\newblock \showarticletitle{${L}_p$ metrics on the {H}eisenberg group and the
  {G}oemans-{L}inial conjecture}. In \bibinfo{booktitle}{{\em Proceedings of
  47th Annual IEEE Symposium on Foundations of Computer Science (FOCS 2006)}}.
  \bibinfo{pages}{99--108}.
\newblock
\newblock
\shownote{Available at
  \url{https://web.math.princeton.edu/~naor/homepage\%20files/L_pHGL.pdf}.}


\bibitem[\protect\citeauthoryear{Lee, Naor, and Peres}{Lee
  et~al\mbox{.}}{2009}]%
        {LNP09}
\bibfield{author}{\bibinfo{person}{James~R. Lee}, \bibinfo{person}{Assaf Naor},
  {and} \bibinfo{person}{Yuval Peres}.} \bibinfo{year}{2009}\natexlab{}.
\newblock \showarticletitle{Trees and {M}arkov convexity}.
\newblock \bibinfo{journal}{{\em Geom. Funct. Anal.\/}} \bibinfo{volume}{18},
  \bibinfo{number}{5} (\bibinfo{year}{2009}), \bibinfo{pages}{1609--1659}.
\newblock
\showCODEN{GFANFB}
\showISSN{1016-443X}
\showDOI{%
\url{http://dx.doi.org/10.1007/s00039-008-0689-0}}


\bibitem[\protect\citeauthoryear{Lee and Sidiropoulos}{Lee and
  Sidiropoulos}{2011}]%
        {LS11}
\bibfield{author}{\bibinfo{person}{James~R. Lee} {and}
  \bibinfo{person}{Anastasios Sidiropoulos}.} \bibinfo{year}{2011}\natexlab{}.
\newblock \showarticletitle{Near-optimal distortion bounds for embedding
  doubling spaces into {$L_1$} [extended abstract]}.
\newblock In \bibinfo{booktitle}{{\em S{TOC}'11---{P}roceedings of the 43rd
  {ACM} {S}ymposium on {T}heory of {C}omputing}}. \bibinfo{publisher}{ACM, New
  York}, \bibinfo{pages}{765--772}.
\newblock
\showDOI{%
\url{http://dx.doi.org/10.1145/1993636.1993737}}


\bibitem[\protect\citeauthoryear{Leighton and Rao}{Leighton and Rao}{1999}]%
        {LR99}
\bibfield{author}{\bibinfo{person}{Tom Leighton} {and} \bibinfo{person}{Satish
  Rao}.} \bibinfo{year}{1999}\natexlab{}.
\newblock \showarticletitle{Multicommodity max-flow min-cut theorems and their
  use in designing approximation algorithms}.
\newblock \bibinfo{journal}{{\em J. ACM\/}} \bibinfo{volume}{46},
  \bibinfo{number}{6} (\bibinfo{year}{1999}), \bibinfo{pages}{787--832}.
\newblock
\showISSN{0004-5411}
\showDOI{%
\url{http://dx.doi.org/10.1145/331524.331526}}


\bibitem[\protect\citeauthoryear{Li}{Li}{2014}]%
        {Li14}
\bibfield{author}{\bibinfo{person}{Sean Li}.} \bibinfo{year}{2014}\natexlab{}.
\newblock \showarticletitle{Coarse differentiation and quantitative
  nonembeddability for {C}arnot groups}.
\newblock \bibinfo{journal}{{\em J. Funct. Anal.\/}} \bibinfo{volume}{266},
  \bibinfo{number}{7} (\bibinfo{year}{2014}), \bibinfo{pages}{4616--4704}.
\newblock
\showISSN{0022-1236}
\showDOI{%
\url{http://dx.doi.org/10.1016/j.jfa.2014.01.026}}


\bibitem[\protect\citeauthoryear{Li}{Li}{2016}]%
        {Li16}
\bibfield{author}{\bibinfo{person}{Sean Li}.} \bibinfo{year}{2016}\natexlab{}.
\newblock \showarticletitle{Markov convexity and nonembeddability of the
  {H}eisenberg group}.
\newblock \bibinfo{journal}{{\em Ann. Inst. Fourier (Grenoble)\/}}
  \bibinfo{volume}{66}, \bibinfo{number}{4} (\bibinfo{year}{2016}),
  \bibinfo{pages}{1615--1651}.
\newblock


\bibitem[\protect\citeauthoryear{Linial}{Linial}{2002a}]%
        {Lin02}
\bibfield{author}{\bibinfo{person}{Nathan Linial}.}
  \bibinfo{year}{2002}\natexlab{a}.
\newblock \showarticletitle{Finite metric-spaces---combinatorics, geometry and
  algorithms}. In \bibinfo{booktitle}{{\em Proceedings of the {I}nternational
  {C}ongress of {M}athematicians, {V}ol. {III} ({B}eijing, 2002)}}.
  \bibinfo{publisher}{Higher Ed. Press}, \bibinfo{address}{Beijing},
  \bibinfo{pages}{573--586}.
\newblock


\bibitem[\protect\citeauthoryear{Linial}{Linial}{2002b}]%
        {Lin-open}
\bibfield{author}{\bibinfo{person}{Nathan Linial}.}
  \bibinfo{year}{2002}\natexlab{b}.
\newblock \showarticletitle{Squared $\ell_2$ metrics into $\ell_1$}.
\newblock In \bibinfo{booktitle}{{\em Open problems on embeddings of finite
  metric spaces, edited by J. Matou\v{s}ek}}. \bibinfo{pages}{5}.
\newblock


\bibitem[\protect\citeauthoryear{Linial, London, and Rabinovich}{Linial
  et~al\mbox{.}}{1995}]%
        {LLR95}
\bibfield{author}{\bibinfo{person}{Nathan Linial}, \bibinfo{person}{Eran
  London}, {and} \bibinfo{person}{Yuri Rabinovich}.}
  \bibinfo{year}{1995}\natexlab{}.
\newblock \showarticletitle{The geometry of graphs and some of its algorithmic
  applications}.
\newblock \bibinfo{journal}{{\em Combinatorica\/}} \bibinfo{volume}{15},
  \bibinfo{number}{2} (\bibinfo{year}{1995}), \bibinfo{pages}{215--245}.
\newblock
\showCODEN{COMBDI}
\showISSN{0209-9683}


\bibitem[\protect\citeauthoryear{Magnani}{Magnani}{2011}]%
        {Mag11}
\bibfield{author}{\bibinfo{person}{Valentino Magnani}.}
  \bibinfo{year}{2011}\natexlab{}.
\newblock \showarticletitle{Area implies coarea}.
\newblock \bibinfo{journal}{{\em Indiana Univ. Math. J.\/}}
  \bibinfo{volume}{60}, \bibinfo{number}{1} (\bibinfo{year}{2011}),
  \bibinfo{pages}{77--100}.
\newblock
\showISSN{0022-2518}
\showDOI{%
\url{http://dx.doi.org/10.1512/iumj.2011.60.4172}}


\bibitem[\protect\citeauthoryear{Makarychev, Makarychev, and
  Vijayaraghavan}{Makarychev et~al\mbox{.}}{2014}]%
        {MMV14}
\bibfield{author}{\bibinfo{person}{Konstantin Makarychev},
  \bibinfo{person}{Yury Makarychev}, {and} \bibinfo{person}{Aravindan
  Vijayaraghavan}.} \bibinfo{year}{2014}\natexlab{}.
\newblock \showarticletitle{Bilu-{L}inial stable instances of max cut and
  minimum multiway cut}. In \bibinfo{booktitle}{{\em Proceedings of the
  {T}wenty-{F}ifth {A}nnual {ACM}-{SIAM} {S}ymposium on {D}iscrete
  {A}lgorithms}}. \bibinfo{publisher}{ACM, New York},
  \bibinfo{pages}{890--906}.
\newblock
\showDOI{%
\url{http://dx.doi.org/10.1137/1.9781611973402.67}}


\bibitem[\protect\citeauthoryear{Mart{\'{\i}}nez, Torrea, and
  Xu}{Mart{\'{\i}}nez et~al\mbox{.}}{2006}]%
        {MTX06}
\bibfield{author}{\bibinfo{person}{Teresa Mart{\'{\i}}nez},
  \bibinfo{person}{Jos{\'e}~L. Torrea}, {and} \bibinfo{person}{Quanhua Xu}.}
  \bibinfo{year}{2006}\natexlab{}.
\newblock \showarticletitle{Vector-valued {L}ittlewood-{P}aley-{S}tein theory
  for semigroups}.
\newblock \bibinfo{journal}{{\em Adv. Math.\/}} \bibinfo{volume}{203},
  \bibinfo{number}{2} (\bibinfo{year}{2006}), \bibinfo{pages}{430--475}.
\newblock
\showCODEN{ADMTA4}
\showISSN{0001-8708}
\showDOI{%
\url{http://dx.doi.org/10.1016/j.aim.2005.04.010}}


\bibitem[\protect\citeauthoryear{Matou{\v{s}}ek}{Matou{\v{s}}ek}{2002a}]%
        {Mat02}
\bibfield{author}{\bibinfo{person}{Ji{\v{r}}{\'{\i}} Matou{\v{s}}ek}.}
  \bibinfo{year}{2002}\natexlab{a}.
\newblock \bibinfo{booktitle}{{\em Lectures on discrete geometry}}.
  \bibinfo{series}{Graduate Texts in Mathematics}, Vol.~\bibinfo{volume}{212}.
\newblock \bibinfo{publisher}{Springer-Verlag, New York}. xvi+481 pages.
\newblock
\showISBNx{0-387-95373-6}
\showDOI{%
\url{http://dx.doi.org/10.1007/978-1-4613-0039-7}}


\bibitem[\protect\citeauthoryear{Matou{\v{s}}ek}{Matou{\v{s}}ek}{2002b}]%
        {Mat02-book}
\bibfield{author}{\bibinfo{person}{Ji{\v{r}}{\'{\i}} Matou{\v{s}}ek}.}
  \bibinfo{year}{2002}\natexlab{b}.
\newblock \bibinfo{booktitle}{{\em Lectures on discrete geometry}}.
  \bibinfo{series}{Graduate Texts in Mathematics}, Vol.~\bibinfo{volume}{212}.
\newblock \bibinfo{publisher}{Springer-Verlag, New York}. xvi+481 pages.
\newblock
\showISBNx{0-387-95373-6}
\showDOI{%
\url{http://dx.doi.org/10.1007/978-1-4613-0039-7}}


\bibitem[\protect\citeauthoryear{Mattila}{Mattila}{1995}]%
        {Mat95}
\bibfield{author}{\bibinfo{person}{Pertti Mattila}.}
  \bibinfo{year}{1995}\natexlab{}.
\newblock \bibinfo{booktitle}{{\em Geometry of sets and measures in {E}uclidean
  spaces}}. \bibinfo{series}{Cambridge Studies in Advanced Mathematics},
  Vol.~\bibinfo{volume}{44}.
\newblock \bibinfo{publisher}{Cambridge University Press, Cambridge}. xii+343
  pages.
\newblock
\showISBNx{0-521-46576-1; 0-521-65595-1}
\showDOI{%
\url{http://dx.doi.org/10.1017/CBO9780511623813}}
\newblock
\shownote{Fractals and rectifiability.}


\bibitem[\protect\citeauthoryear{Mendel and Naor}{Mendel and Naor}{2013}]%
        {MN13}
\bibfield{author}{\bibinfo{person}{Manor Mendel} {and} \bibinfo{person}{Assaf
  Naor}.} \bibinfo{year}{2013}\natexlab{}.
\newblock \showarticletitle{Markov convexity and local rigidity of distorted
  metrics}.
\newblock \bibinfo{journal}{{\em J. Eur. Math. Soc. (JEMS)\/}}
  \bibinfo{volume}{15}, \bibinfo{number}{1} (\bibinfo{year}{2013}),
  \bibinfo{pages}{287--337}.
\newblock
\showISSN{1435-9855}
\showDOI{%
\url{http://dx.doi.org/10.4171/JEMS/362}}


\bibitem[\protect\citeauthoryear{Milman and Wolfson}{Milman and
  Wolfson}{1978}]%
        {MW78}
\bibfield{author}{\bibinfo{person}{V.~D. Milman} {and} \bibinfo{person}{H.
  Wolfson}.} \bibinfo{year}{1978}\natexlab{}.
\newblock \showarticletitle{Minkowski spaces with extremal distance from the
  {E}uclidean space}.
\newblock \bibinfo{journal}{{\em Israel J. Math.\/}} \bibinfo{volume}{29},
  \bibinfo{number}{2-3} (\bibinfo{year}{1978}), \bibinfo{pages}{113--131}.
\newblock
\showISSN{0021-2172}


\bibitem[\protect\citeauthoryear{Naor}{Naor}{2010}]%
        {Nao10}
\bibfield{author}{\bibinfo{person}{Assaf Naor}.}
  \bibinfo{year}{2010}\natexlab{}.
\newblock \showarticletitle{{$L_1$} embeddings of the {H}eisenberg group and
  fast estimation of graph isoperimetry}. In \bibinfo{booktitle}{{\em
  Proceedings of the {I}nternational {C}ongress of {M}athematicians. {V}olume
  {III}}}. \bibinfo{publisher}{Hindustan Book Agency, New Delhi},
  \bibinfo{pages}{1549--1575}.
\newblock


\bibitem[\protect\citeauthoryear{Naor}{Naor}{2012}]%
        {Nao12}
\bibfield{author}{\bibinfo{person}{Assaf Naor}.}
  \bibinfo{year}{2012}\natexlab{}.
\newblock \showarticletitle{An introduction to the {R}ibe program}.
\newblock \bibinfo{journal}{{\em Jpn. J. Math.\/}} \bibinfo{volume}{7},
  \bibinfo{number}{2} (\bibinfo{year}{2012}), \bibinfo{pages}{167--233}.
\newblock
\showISSN{0289-2316}
\showDOI{%
\url{http://dx.doi.org/10.1007/s11537-012-1222-7}}


\bibitem[\protect\citeauthoryear{Naor}{Naor}{2014}]%
        {Nao14}
\bibfield{author}{\bibinfo{person}{Assaf Naor}.}
  \bibinfo{year}{2014}\natexlab{}.
\newblock \showarticletitle{Comparison of metric spectral gaps}.
\newblock \bibinfo{journal}{{\em Anal. Geom. Metr. Spaces\/}}
  \bibinfo{volume}{2} (\bibinfo{year}{2014}), \bibinfo{pages}{1--52}.
\newblock
\showISSN{2299-3274}
\showDOI{%
\url{http://dx.doi.org/10.2478/agms-2014-0001}}


\bibitem[\protect\citeauthoryear{Naor, Peres, Schramm, and Sheffield}{Naor
  et~al\mbox{.}}{2006}]%
        {NPSS06}
\bibfield{author}{\bibinfo{person}{Assaf Naor}, \bibinfo{person}{Yuval Peres},
  \bibinfo{person}{Oded Schramm}, {and} \bibinfo{person}{Scott Sheffield}.}
  \bibinfo{year}{2006}\natexlab{}.
\newblock \showarticletitle{Markov chains in smooth {B}anach spaces and
  {G}romov-hyperbolic metric spaces}.
\newblock \bibinfo{journal}{{\em Duke Math. J.\/}} \bibinfo{volume}{134},
  \bibinfo{number}{1} (\bibinfo{year}{2006}), \bibinfo{pages}{165--197}.
\newblock
\showCODEN{DUMJAO}
\showISSN{0012-7094}
\showDOI{%
\url{http://dx.doi.org/10.1215/S0012-7094-06-13415-4}}


\bibitem[\protect\citeauthoryear{Naor, Rabani, and Sinclair}{Naor
  et~al\mbox{.}}{2005}]%
        {NRS05}
\bibfield{author}{\bibinfo{person}{Assaf Naor}, \bibinfo{person}{Yuval Rabani},
  {and} \bibinfo{person}{Alistair Sinclair}.} \bibinfo{year}{2005}\natexlab{}.
\newblock \showarticletitle{Quasisymmetric embeddings, the observable diameter,
  and expansion properties of graphs}.
\newblock \bibinfo{journal}{{\em J. Funct. Anal.\/}} \bibinfo{volume}{227},
  \bibinfo{number}{2} (\bibinfo{year}{2005}), \bibinfo{pages}{273--303}.
\newblock
\showCODEN{JFUAAW}
\showISSN{0022-1236}
\showDOI{%
\url{http://dx.doi.org/10.1016/j.jfa.2005.04.003}}


\bibitem[\protect\citeauthoryear{Naor and Silberman}{Naor and
  Silberman}{2011}]%
        {NS11}
\bibfield{author}{\bibinfo{person}{Assaf Naor} {and} \bibinfo{person}{Lior
  Silberman}.} \bibinfo{year}{2011}\natexlab{}.
\newblock \showarticletitle{Poincar\'e inequalities, embeddings, and wild
  groups}.
\newblock \bibinfo{journal}{{\em Compos. Math.\/}} \bibinfo{volume}{147},
  \bibinfo{number}{5} (\bibinfo{year}{2011}), \bibinfo{pages}{1546--1572}.
\newblock
\showISSN{0010-437X}
\showDOI{%
\url{http://dx.doi.org/10.1112/S0010437X11005343}}


\bibitem[\protect\citeauthoryear{Naor and Young}{Naor and Young}{2017}]%
        {NY-full}
\bibfield{author}{\bibinfo{person}{A. Naor} {and} \bibinfo{person}{R. Young}.}
  \bibinfo{year}{2017}\natexlab{}.
\newblock \bibinfo{title}{Vertical perimeter versus horizontal perimeter}.
  (\bibinfo{year}{2017}).
\newblock
\newblock
\shownote{Preprint available at \url{https://arxiv.org/abs/1701.00620}.}


\bibitem[\protect\citeauthoryear{Ostrovskii}{Ostrovskii}{2013}]%
        {Ost13}
\bibfield{author}{\bibinfo{person}{Mikhail~I. Ostrovskii}.}
  \bibinfo{year}{2013}\natexlab{}.
\newblock \bibinfo{booktitle}{{\em Metric embeddings}}. \bibinfo{series}{De
  Gruyter Studies in Mathematics}, Vol.~\bibinfo{volume}{49}.
\newblock \bibinfo{publisher}{De Gruyter, Berlin}. xii+372 pages.
\newblock
\showISBNx{978-3-11-026340-4; 978-3-11-026401-2}
\showDOI{%
\url{http://dx.doi.org/10.1515/9783110264012}}
\newblock
\shownote{Bilipschitz and coarse embeddings into Banach spaces.}


\bibitem[\protect\citeauthoryear{Pansu}{Pansu}{1982}]%
        {Pan82}
\bibfield{author}{\bibinfo{person}{Pierre Pansu}.}
  \bibinfo{year}{1982}\natexlab{}.
\newblock \showarticletitle{Une in\'egalit\'e isop\'erim\'etrique sur le groupe
  de {H}eisenberg}.
\newblock \bibinfo{journal}{{\em C. R. Acad. Sci. Paris S\'er. I Math.\/}}
  \bibinfo{volume}{295}, \bibinfo{number}{2} (\bibinfo{year}{1982}),
  \bibinfo{pages}{127--130}.
\newblock
\showCODEN{CASMEI}
\showISSN{0249-6291}


\bibitem[\protect\citeauthoryear{Pansu}{Pansu}{1989}]%
        {Pan89}
\bibfield{author}{\bibinfo{person}{Pierre Pansu}.}
  \bibinfo{year}{1989}\natexlab{}.
\newblock \showarticletitle{M\'etriques de {C}arnot-{C}arath\'eodory et
  quasiisom\'etries des espaces sym\'etriques de rang un}.
\newblock \bibinfo{journal}{{\em Ann. of Math. (2)\/}} \bibinfo{volume}{129},
  \bibinfo{number}{1} (\bibinfo{year}{1989}), \bibinfo{pages}{1--60}.
\newblock
\showCODEN{ANMAAH}
\showISSN{0003-486X}
\showDOI{%
\url{http://dx.doi.org/10.2307/1971484}}


\bibitem[\protect\citeauthoryear{Rao}{Rao}{1999}]%
        {Rao99}
\bibfield{author}{\bibinfo{person}{Satish Rao}.}
  \bibinfo{year}{1999}\natexlab{}.
\newblock \showarticletitle{Small distortion and volume preserving embeddings
  for planar and {E}uclidean metrics}. In \bibinfo{booktitle}{{\em Proceedings
  of the {F}ifteenth {A}nnual {S}ymposium on {C}omputational {G}eometry
  ({M}iami {B}each, {FL}, 1999)}}. \bibinfo{publisher}{ACM, New York},
  \bibinfo{pages}{300--306 (electronic)}.
\newblock
\showDOI{%
\url{http://dx.doi.org/10.1145/304893.304983}}


\bibitem[\protect\citeauthoryear{Shahrokhi and Matula}{Shahrokhi and
  Matula}{1990}]%
        {SM90}
\bibfield{author}{\bibinfo{person}{Farhad Shahrokhi} {and}
  \bibinfo{person}{D.~W. Matula}.} \bibinfo{year}{1990}\natexlab{}.
\newblock \showarticletitle{The maximum concurrent flow problem}.
\newblock \bibinfo{journal}{{\em J. Assoc. Comput. Mach.\/}}
  \bibinfo{volume}{37}, \bibinfo{number}{2} (\bibinfo{year}{1990}),
  \bibinfo{pages}{318--334}.
\newblock
\showCODEN{JACOAH}
\showISSN{0004-5411}


\bibitem[\protect\citeauthoryear{Shmoys}{Shmoys}{1997}]%
        {Shm95}
\bibfield{author}{\bibinfo{person}{D.~B. Shmoys}.}
  \bibinfo{year}{1997}\natexlab{}.
\newblock \showarticletitle{Cut problems and their application to
  divide-and-conquer}.
\newblock In \bibinfo{booktitle}{{\em Approximation Algorithms for NP-hard
  Problems, (D.S. Hochbaum, ed.)}}. \bibinfo{publisher}{PWS},
  \bibinfo{pages}{192--235}.
\newblock


\bibitem[\protect\citeauthoryear{Sinclair and Jerrum}{Sinclair and
  Jerrum}{1989}]%
        {JS89}
\bibfield{author}{\bibinfo{person}{Alistair Sinclair} {and}
  \bibinfo{person}{Mark Jerrum}.} \bibinfo{year}{1989}\natexlab{}.
\newblock \showarticletitle{Approximate counting, uniform generation and
  rapidly mixing {M}arkov chains}.
\newblock \bibinfo{journal}{{\em Inform. and Comput.\/}} \bibinfo{volume}{82},
  \bibinfo{number}{1} (\bibinfo{year}{1989}), \bibinfo{pages}{93--133}.
\newblock
\showISSN{0890-5401}
\showDOI{%
\url{http://dx.doi.org/10.1016/0890-5401(89)90067-9}}


\bibitem[\protect\citeauthoryear{Tessera}{Tessera}{2008}]%
        {Tes08}
\bibfield{author}{\bibinfo{person}{Romain Tessera}.}
  \bibinfo{year}{2008}\natexlab{}.
\newblock \showarticletitle{Quantitative property {A}, {P}oincar\'e
  inequalities, {$L^p$}-compression and {$L^p$}-distortion for metric measure
  spaces}.
\newblock \bibinfo{journal}{{\em Geom. Dedicata\/}}  \bibinfo{volume}{136}
  (\bibinfo{year}{2008}), \bibinfo{pages}{203--220}.
\newblock
\showCODEN{GEMDAT}
\showISSN{0046-5755}
\showDOI{%
\url{http://dx.doi.org/10.1007/s10711-008-9286-5}}


\bibitem[\protect\citeauthoryear{Trevisan}{Trevisan}{2012}]%
        {Tre12}
\bibfield{author}{\bibinfo{person}{Luca Trevisan}.}
  \bibinfo{year}{2012}\natexlab{}.
\newblock \showarticletitle{On {K}hot's unique games conjecture}.
\newblock \bibinfo{journal}{{\em Bull. Amer. Math. Soc. (N.S.)\/}}
  \bibinfo{volume}{49}, \bibinfo{number}{1} (\bibinfo{year}{2012}),
  \bibinfo{pages}{91--111}.
\newblock
\showCODEN{BAMOAD}
\showISSN{0273-0979}
\showDOI{%
\url{http://dx.doi.org/10.1090/S0273-0979-2011-01361-1}}


\bibitem[\protect\citeauthoryear{Witsenhausen}{Witsenhausen}{1986}]%
        {Wit86}
\bibfield{author}{\bibinfo{person}{H.~S. Witsenhausen}.}
  \bibinfo{year}{1986}\natexlab{}.
\newblock \showarticletitle{Minimum dimension embedding of finite metric
  spaces}.
\newblock \bibinfo{journal}{{\em J. Combin. Theory Ser. A\/}}
  \bibinfo{volume}{42}, \bibinfo{number}{2} (\bibinfo{year}{1986}),
  \bibinfo{pages}{184--199}.
\newblock
\showCODEN{JCBTA7}
\showISSN{0097-3165}
\showDOI{%
\url{http://dx.doi.org/10.1016/0097-3165(86)90089-0}}


\end{thebibliography}

\end{document}